# Spectroscopic Study of M33 with the LAMOST Survey. I. Chemical Gradients from Nebulae

Sofya Alexeeva[1] and Gang Zhao[1,2]
[1] CAS Key Laboratory of Optical Astronomy, National Astronomical Observatories, Chinese Academy of Sciences, Beijing, 100101, People's Republic of China alexeeva@nao.cas.cn
[2] School of Astronomy and Space Science, University of Chinese Academy of Sciences, Beijing, 100049, People's Republic of China; gzhao@nao.cas.cn

## Abstract

Morphological and chemical structures of M33 are investigated with the LAMOST DR7 survey. Physical parameters; extinction; chemical composition of He, N, O, Ne, S, Cl, and Ar (where available); and radial velocities were determined for 110 nebulae (95 H II regions and 15 planetary nebulae) in M33. Among them, 8 planetary nebulae and 55 H II regions in M33 are newly discovered. We obtained the following O abundance gradients: $-0.199^{+0.030}_{-0.030}$ dex $R_{25}^{-1}$ (based on 95 H II regions), $-0.124^{+0.036}_{-0.036}$ dex $R_{25}^{-1}$ (based on 93 H II regions), and $-0.207^{+0.160}_{-0.174}$ dex $R_{25}^{-1}$ (based on 21 H II regions), utilizing abundances from N2 at O3N2 diagnostics and the $T_e$-sensitive method, respectively. The He, N, Ne, S, and Ar gradients resulted in slopes of $-0.179^{+0.145}_{-0.146}$, $-0.431^{+0.282}_{-0.281}$, $-0.171^{+0.234}_{-0.239}$, $-0.417^{+0.174}_{-0.182}$, and $-0.340^{+0.156}_{-0.157}$, respectively, utilizing abundances from the $T_e$-sensitive method. Our results confirm the existence of the negative axisymmetric global metallicity distribution that is assumed in the literature. We noticed one new WC star candidate and one transition W-R WN/C candidate. The grand-design pattern of the spiral structure of M33 is presented.

*Key words:* Gaseous nebulae – Triangulum Galaxy – Chemical abundances – Interstellar abundances – Galaxy chemical evolution – Interstellar medium – Emission nebulae – Nebulae

## 1. Introduction

M33 (NGC 598, Triangulum galaxy) is the third-largest member of our Local Group of galaxies following the Andromeda galaxy (M31) and the Milky Way. Spectroscopic studies of M33 started a long time ago, when the bright lines in the nebulae were observed by Slipher (1915), Pease (1915), and Hubble (1926). Almost 100 yr ago Edwin Hubble noticed that the great angular diameter of M33 and its high degree of resolution provide an exceptional opportunity for detailed investigation. Therefore, M33 is still considered by astronomers as a perfect astrophysical laboratory for studying star formation processes, stellar evolution, galacto-chemical evolution, and the properties of the interstellar medium.

The distribution of chemical abundances in M33 and other galaxies was originally investigated by Searle (1971). His discovery of a negative metallicity gradient outward in spiral galaxies subsequently has been confirmed by numerous studies (e.g., Pagel et al. 1979; Kwitter & Aller 1981; Garnett & Shields 1987). The metallicity gradients have engaged much attention, because they reflect the evolution of the galaxy and can provide better understanding of complex processes, such as star formation, stellar migration, gas flows within the disk, radial gas flows, and metal-enriched gas outflows.

Chemical gradients are commonly tracked by measuring the abundance of metals such as O, Ne, S, and Ar as a function of galactocentric distance. Traditionally, oxygen is the most common element for determination of the abundance gradient. Oxygen abundances are well determined from ionic abundances of $O^+$ and $O^{++}$, whose lines are observed in the visual range of wavelengths. However, it should be noted that the oxygen abundance measured from the spectral lines of planetary nebulae (PNe) could differ from the oxygen abundance of the preexisting gas out of which the progenitor stars originated. Oxygen could be produced or destroyed in the stellar progenitors. Delgado-Inglada et al. (2015) presented evidence that oxygen can be produced at low metallicity, and their analysis showed that argon and chlorine are the best metallicity indicators of the progenitors of PNe.

There are many studies where metallicity gradients were derived from H II regions, PNe, and massive supergiant stars in M33, e.g., Monteverde et al. (1997), Urbaneja et al. (2005), Crockett et al. (2006), U et al. (2009), Magrini et al. (2009), Rosolowsky & Simon (2008), Magrini et al. (2010), Bresolin et al. (2010), Bresolin (2011), Toribio San Cipriano et al. (2016), Lin et al. (2017), and others. The best review of current observational knowledge concerning the abundances of the ionized gas (H II regions and PNe) in nearby galaxies, including M33, and how they inform us about the time evolution of metallicity gradients can be found in Bresolin (2015).

Magrini et al. (2009, 2010) claimed that H II regions and PNe have oxygen gradients very close within the errors. Their result was confirmed by Bresolin et al. (2010), who found no significant oxygen abundance offset between PNe and H II regions at any given galactocentric distance despite their different age groups. It would be reasonable to expect that the gradients obtained using H II regions and massive stars are the same, since the two types of objects represent the same young galactic population. Such a kind of comparison has been carried out in M33, and a difference on the order of 0.2 dex was reported in Bresolin (2015). The origin of this disagreement is still not clear. Peña & Flores-Durán (2019) analyzed galactocentric radii and chemical abundances from the literature and inferred that the metallicity gradients shown by PNe are flatter than those of H II regions.

The theoretical models of galacto-chemical evolution can predict both directions of gradient evolution: the gradient could become steeper or flatter with time (Kang et al. 2012; Mollá et al. 2019). There are many factors playing a role, such as gas infall, gas outflow, and turbulence in the interstellar medium. All these are based on observations of H II regions, PNe, and massive stars representing different age groups in the evolution of the galaxy. To infer a reliable chemical history of M33, a larger and more homogeneous sample of H II regions, PNe, and B-type stars is demanded in order to get observational constraints on the real gradients.

M33 is a spiral galaxy with SA(s)cd Hubble classification. Earlier studies of the structure of M33 are based on optical images, e.g., Boulesteix et al. (1974), Sandage & Humphreys (1980), where Sandage & Humphreys (1980) identified 10 spiral arms, five on each side of the center. In the optical bands, M33 could be considered as a flocculent spiral galaxy, while UV and IR observations show more prominent arms. The spiral structure of M33 has also been examined on the basis of 197 OB associations by Ivanov & Kunchev (1985). They pointed out a system of seven spiral arms. Although there are many recent studies about the spiral structure of M33 (e.g., Li & Henning 2011; Kam et al. 2015), the arms are traced slightly differently in various sources, and M33 does not have a clearly defined grand-design spiral pattern.

In this paper we present a spectroscopic study of M33 with spectra obtained with the Guoshoujing Telescope (the Large Sky Area Multi-Object Fiber Spectroscopic Telescope, LAMOST) from 2011 to 2018 yr. A large and homogeneous sample of 110 nebulae (95 H II regions and 15 PNe) that spreads over the whole M33 disk is investigated. More than a half of all nebulae are investigated for the fist time. The spectra are analyzed to determine abundances of the elements He, N, O, Ne, S, Cl, and Ar represented by observed emission lines. Different methods for calculating chemical abundances in photoionized nebulae are employed: the direct method based on the analysis of collisionally excited lines (CELs), and the method of strong-line abundances with O3N2 and N2 diagnostics and optical recombination lines (ORLs). We hope that our study will help to shed light on the evolution of the gradient of oxygen and other elements in M33.

The article is organized as follows. Section 2 describes the observations, reduction techniques, and target selection. The method of calculations is presented in Section 3. The radial velocities of the sample nebula are calculated in Section 4.2. In Section 4, we investigate morphological and chemical structures of M33. Comparisons with the previous studies are presented in Section 5. We summarize our conclusions in Section 6.

## 2. Observations, Target Selection, and Classification

The Large sky Area Multi-Object Spectroscopic Telescope (LAMOST) is a 4.0 m telescope located at the Xinglong Observatory northeast of Beijing, China. The LAMOST survey is a spectroscopic survey that eventually covers approximately half of the celestial sphere and collects 10 million spectra of stars, galaxies, and quasars (Zhao et al. 2006, 2012; Cui et al. 2012).

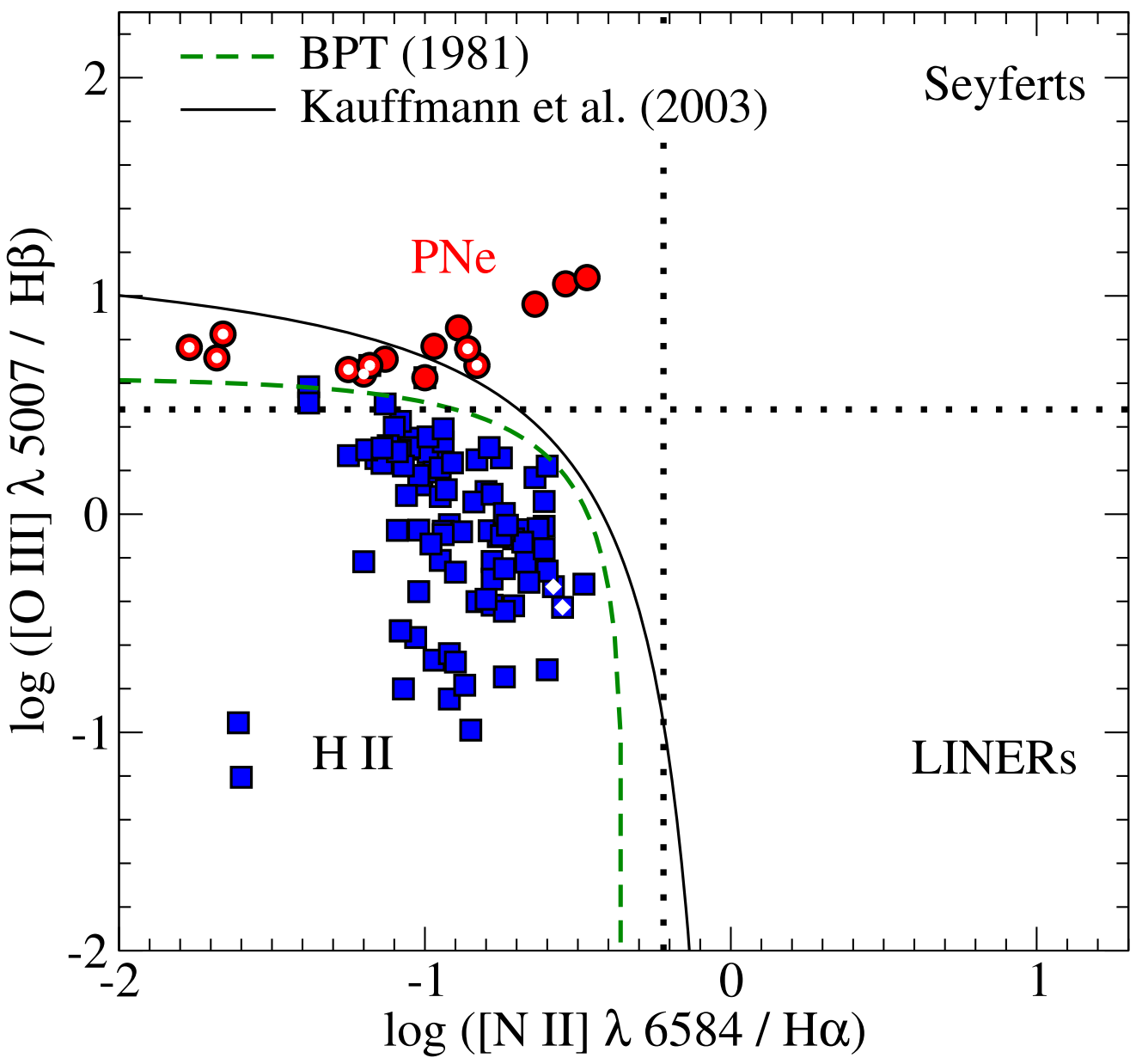

**Figure 1.** BPT diagram with 15 PNe and 95 H II regions. The solid curve is a demarcation line between starbursts and AGNs according to Kauffmann et al. (2003), the dashed curve is a demarcation line between PNe and H II regions according to Baldwin et al. (1981), and dotted lines are demarcations between LINERs and Seyferts. PNe are shown by circles, among which new PNe are marked by white central circles and H II regions are shown by squares. The H II regions J013402.76+303833.2 (#26) and J013350.88+303712.2 (#37) are marked by white central diamonds.

We checked 2154 LAMOST low-resolution ($R \sim 1800$) spectra at 3700–9099 Å from the DR7 release (observed from 2011 December to 2018 October) in the radius of 31′.6 from the center of the M33. A total of 110 high signal-to-noise ratio nebular spectra were selected for further analysis. They were divided into two groups: H II regions and PNe. Objects with emission-line spectra were classified with the commonly used Baldwin–Phillips–Terlevich (BPT) criteria from Baldwin et al. (1981) to distinguish the PNe and H II regions, and criteria from Kauffmann et al. (2003) to distinguish the starburst and active galactic nuclei (AGNs). It should be noted that other classification dividers are available in the literature (e.g., Kniazev et al. 2008; Sanders et al. 2012). The BPT diagram with our objects is presented in Figure 1.

We have identified 95 H II regions and 15 PNe. The number of H II regions is higher because H II regions are much larger (~10–30 pc) than PNe (~0.1 pc), so the probability of finding H II is higher during the survey. The whole sample of PNe and H II regions is presented in Table 1, and their distribution in M33 is plotted in Figure 2. We call regions “new” if there is no previous classification as an H II region or PN in the literature for objects within the radius of 2″ (the site seeing of the telescope). For regions that were classified in previous works, we provide references in Table 1.

Recently, Zhang et al. (2020a) presented a catalog of H$\alpha$ emission-line point sources in the vicinity of M31 and M33 observed with LAMOST at the period from 2011 September to 2016 January. For classification, Zhang et al. (2020a) adopted the commonly used criteria from Baldwin et al. (1981) to distinguish the PNe and H II regions and criteria from Sabbadin et al. (1977), Canto (1981), and Riesgo & López (2006) to separate the supernova remnants (SNRs) from PNe and H II

**Table 1**
Sample of PNe and H II Regions in M33 from LAMOST Survey

| No. | Name LAMOST | Date YY-MM-DD | RA J2000 | Dec J2000 | New | No. | Name LAMOST | Date YY-MM-DD | R.A. J2000 | Decl. J2000 | New |
|---|---|---|---|---|---|---|---|---|---|---|---|
| (1) | (2) | (3) | (4) | (5) | (6) | (7) | (8) | (9) | (10) | (11) | (12) |
| | | PNe | | | | 55 | J013347.05+302723.3 | 2013-10-04 | 23.446061 | 30.456476 | CPS87 |
| 1 | J013311.44+304856.8 | 2011-12-11 | 23.297667 | 30.815805 | MOH2006 | 56 | J013355.22+303430.2 | 2013-10-04 | 23.480087 | 30.575057 | LJD2011 |
| 2 | J013413.33+304501.1 | 2011-12-17 | 23.555580 | 30.750330 | CDL2004 | 57 | J013425.41+305257.2 | 2013-12-23 | 23.605916 | 30.882556 | new |
| 3 | J013316.06+305643.2 | 2011-12-17 | 23.316958 | 30.945360 | new | 58 | J013414.54+303427.0 | 2013-12-23 | 23.560624 | 30.574168 | new |
| 4 | J013332.34+304240.3 | 2011-12-17 | 23.384750 | 30.711220 | ZCH2020 | 59 | J013439.45+303114.3 | 2013-12-23 | 23.664416 | 30.520639 | new |
| 5 | J013401.14+305027.0 | 2011-12-22 | 23.504750 | 30.840861 | CDL2004 | 60 | J013339.57+302140.4 | 2013-12-23 | 23.414916 | 30.361223 | new |
| 6 | J013302.90+301122.5 | 2011-12-25 | 23.262117 | 30.189594 | new | 61 | J013353.56+303322.8 | 2013-12-23 | 23.473208 | 30.556360 | new |
| 7 | J013231.93+303516.7 | 2011-12-25 | 23.133083 | 30.587973 | new | 62 | J013418.22+302202.6 | 2013-12-23 | 23.575958 | 30.367389 | CS82 |
| 8 | J013316.19+305245.4 | 2013-09-25 | 23.317479 | 30.879290 | new | 63 | J013410.54+302556.7 | 2013-12-23 | 23.543957 | 30.432444 | LHK2017 |
| 9 | J013311.39+304856.5 | 2013-09-25 | 23.297489 | 30.815715 | MOH2006 | 64 | J013252.83+303340.0 | 2013-12-23 | 23.220166 | 30.561138 | new |
| 10 | J013316.51+305249.4 | 2013-09-25 | 23.318810 | 30.880406 | new | 65 | J013301.67+303027.9 | 2013-12-23 | 23.256999 | 30.507750 | ZCH2020 |
| 11 | J013302.79+302326.8 | 2013-12-23 | 23.261666 | 30.390805 | new | 66 | J013413.62+304829.8 | 2013-12-23 | 23.556791 | 30.808305 | new |
| 12 | J013416.39+304326.5 | 2014-12-24 | 23.568292 | 30.724028 | MCM2000 | 67 | J013437.36+303454.2 | 2013-12-23 | 23.655707 | 30.581723 | BCL74 |
| 13 | J013350.84+305639.0 | 2014-12-24 | 23.461874 | 30.944194 | new | 68 | J013417.11+303426.7 | 2014-01-01 | 23.571332 | 30.574084 | new |
| 14 | J013315.14+305645.6 | 2014-12-24 | 23.313124 | 30.946027 | new | 69 | J013256.52+303415.1 | 2014-01-01 | 23.235541 | 30.570862 | new |
| 15 | J013302.02+301108.2 | 2014-12-24 | 23.258417 | 30.185612 | ZCH2020 | 70 | J013355.21+303403.3 | 2014-01-01 | 23.480083 | 30.567584 | new |
| | | H II regions | | | | 71 | J013244.05+302206.0 | 2014-01-01 | 23.183583 | 30.368361 | new |
| 16 | J013406.62+304147.7 | 2011-12-11 | 23.527624 | 30.696611 | VGHC86 | 72 | J013307.37+302314.2 | 2014-01-01 | 23.280749 | 30.387305 | new |
| 17 | J013334.27+304136.7 | 2011-12-11 | 23.392792 | 30.693528 | DMS93 | 73 | J013450.09+305446.4 | 2014-01-02 | 23.708749 | 30.912916 | new |
| 18 | J013334.86+303705.5 | 2011-12-11 | 23.395252 | 30.618199 | BCL74 | 74 | J013359.00+304910.8 | 2014-01-02 | 23.495874 | 30.819668 | BCL74 |
| 19 | J013303.09+303101.8 | 2011-12-13 | 23.262875 | 30.517167 | new | 75 | J013333.03+304206.3 | 2014-01-02 | 23.387666 | 30.701751 | new |
| 20 | J013346.12+303653.7 | 2011-12-13 | 23.442208 | 30.614944 | new | 76 | J013446.56+303814.2 | 2014-01-02 | 23.694041 | 30.637278 | new |
| 21 | J013439.00+310231.0 | 2011-12-17 | 23.662500 | 31.041945 | ZCH2020 | 77 | J013357.56+304218.2 | 2014-01-02 | 23.489874 | 30.705056 | new |
| 22 | J013307.33+302314.2 | 2011-12-17 | 23.280542 | 30.387297 | new | 78 | J013442.53+305544.3 | 2014-12-24 | 23.677249 | 30.928999 | new |
| 23 | J013441.32+302811.6 | 2011-12-17 | 23.672167 | 30.469900 | LHK2017 | 79 | J013422.21+304702.7 | 2014-12-24 | 23.592582 | 30.784111 | new |
| 24 | J013334.22+304127.6 | 2011-12-17 | 23.392624 | 30.691027 | new | 80 | J013341.97+304620.1 | 2014-12-24 | 23.424908 | 30.772254 | new |
| 25 | J013459.43+303701.6 | 2011-12-17 | 23.747643 | 30.617119 | new | 81 | J013330.18+305023.2 | 2014-12-24 | 23.375773 | 30.839784 | new |
| 26 | J013402.76+303833.2 | 2011-12-17 | 23.511540 | 30.642580 | CDL2004 | 82 | J013313.57+302236.3 | 2014-12-24 | 23.306583 | 30.376750 | ZCH2020 |
| 27 | J013306.52+303010.8 | 2011-12-17 | 23.277208 | 30.503000 | new | 83 | J013256.76+303417.1 | 2014-12-24 | 23.236500 | 30.571444 | new |
| 28 | J013348.14+303928.6 | 2011-12-17 | 23.450624 | 30.657972 | new | 84 | J013340.36+302001.6 | 2014-12-24 | 23.418208 | 30.333778 | (SNR?) ZCH2020 |
| 29 | J013439.70+304406.5 | 2011-12-22 | 23.665458 | 30.735144 | new | 85 | J013442.54+305544.5 | 2014-12-24 | 23.677250 | 30.929028 | ZCH2020 |
| 30 | J013313.47+304527.1 | 2011-12-22 | 23.306125 | 30.757528 | new | 86 | J013500.29+304150.8 | 2014-12-24 | 23.751249 | 30.697471 | LHK2017 |
| 31 | J013344.10+304836.2 | 2011-12-22 | 23.433750 | 30.810056 | new | 87 | J013505.78+304101.4 | 2014-12-24 | 23.774110 | 30.683732 | LHK2017 |
| 32 | J013316.36+305320.5 | 2011-12-22 | 23.318167 | 30.889028 | new | 88 | J013413.71+310401.4 | 2014-12-24 | 23.557136 | 31.067058 | new |
| 33 | J013344.56+303201.2 | 2011-12-22 | 23.435667 | 30.533694 | B2011 | 89 | J013336.55+305035.5 | 2014-12-24 | 23.402292 | 30.843208 | LHK2017 |
| 34 | J013346.95+302722.6 | 2011-12-22 | 23.445625 | 30.456303 | MVM2007 | 90 | J013404.06+304658.2 | 2014-12-24 | 23.516957 | 30.782861 | LHK2017 |
| 35 | J013358.66+303526.0 | 2011-12-22 | 23.494458 | 30.590581 | BCL74 | 91 | J013249.87+304555.4 | 2014-12-24 | 23.207833 | 30.765389 | new |
| 36 | J013231.96+303457.7 | 2011-12-22 | 23.133167 | 30.582722 | new | 92 | J013309.23+302330.4 | 2014-12-24 | 23.288499 | 30.391805 | new |
| 37 | J013350.88+303712.2 | 2011-12-22 | 23.462000 | 30.620083 | MCM2000 | 93 | J013447.26+303755.8 | 2014-12-24 | 23.696958 | 30.632167 | new |
| 38 | J013254.06+302321.4 | 2011-12-22 | 23.225250 | 30.389278 | RVP2013 | 94 | J013359.51+303442.8 | 2014-12-24 | 23.497999 | 30.578556 | new |
| 39 | J013309.64+302325.5 | 2011-12-25 | 23.290208 | 30.390425 | BCL74 | 95 | J013259.43+303445.9 | 2014-12-24 | 23.247625 | 30.579418 | LHK2017 |
| 40 | J013345.00+302138.0 | 2011-12-25 | 23.437500 | 30.360564 | new | 96 | J013244.13+303855.2 | 2014-12-24 | 23.183916 | 30.648694 | new |
| 41 | J013351.46+304057.0 | 2011-12-25 | 23.464417 | 30.682500 | new | 97 | J013254.62+302320.6 | 2014-12-24 | 23.227620 | 30.389060 | new |
| 42 | J013307.63+303100.6 | 2011-12-25 | 23.281833 | 30.516861 | new | 98 | J013238.79+304109.0 | 2014-12-24 | 23.161666 | 30.685860 | new |
| 43 | J013244.58+303459.2 | 2011-12-25 | 23.185750 | 30.583122 | BCL74 | 99 | J013335.28+304146.0 | 2014-12-24 | 23.397025 | 30.696127 | new |
| 44 | J013354.61+303308.2 | 2012-01-13 | 23.477566 | 30.552284 | new | 100 | J013343.99+310210.6 | 2018-10-09 | 23.433310 | 31.036300 | new |

**Table 1**
(Continued)

| No. (1) | Name LAMOST (2) | Date YY-MM-DD (3) | RA J2000 (4) | Dec J2000 (5) | New (6) | No. (7) | Name LAMOST (8) | Date YY-MM-DD (9) | R.A. J2000 (10) | Decl. J2000 (11) | New (12) |
|---|---|---|---|---|---|---|---|---|---|---|---|
| 45 | J013334.86+303705.4 | 2012-01-13 | 23.395279 | 30.618169 | BCL74 | 101 | J013415.50+303711.5 | 2018-10-09 | 23.564621 | 30.619886 | LHK2017 |
| 46 | J013432.29+304659.3 | 2013-09-25 | 23.634559 | 30.783161 | LHK2017 | 102 | J013339.23+303804.3 | 2018-10-09 | 23.413473 | 30.634550 | new |
| 47 | J013342.95+304440.6 | 2013-09-25 | 23.428971 | 30.744614 | BCL74 | 103 | J013339.48+304540.5 | 2018-10-09 | 23.414521 | 30.761270 | new |
| 48 | J013311.22+304515.3 | 2013-09-25 | 23.296773 | 30.754255 | VGHC86 | 104 | J013431.65+304717.5 | 2018-10-09 | 23.631912 | 30.788216 | MOH2006 |
| 49 | J013406.80+304725.8 | 2013-09-25 | 23.528339 | 30.790500 | BCL74 | 105 | J013505.71+304101.6 | 2018-10-09 | 23.773820 | 30.683798 | LHK2017 |
| 50 | J013433.41+304653.8 | 2013-09-25 | 23.639241 | 30.781620 | new | 106 | J013357.70+301714.1 | 2018-10-09 | 23.490420 | 30.287250 | LHK2017 |
| 51 | J013354.54+303310.0 | 2013-10-04 | 23.477259 | 30.552790 | new | 107 | J013256.31+302733.2 | 2018-10-09 | 23.234634 | 30.459227 | new |
| 52 | J013310.67+302714.6 | 2013-10-04 | 23.294479 | 30.454074 | new | 108 | J013439.41+303114.0 | 2018-10-09 | 23.664231 | 30.520579 | new |
| 53 | J013324.70+303049.8 | 2013-10-04 | 23.352922 | 30.513834 | RVP2013 | 109 | J013307.47+304258.3 | 2018-10-09 | 23.281158 | 30.716221 | new |
| 54 | J013301.32+303044.1 | 2013-10-04 | 23.255527 | 30.512264 | ZCH2020 | 110 | J013237.67+304005.4 | 2018-10-09 | 23.156984 | 30.668187 | LHK2017 |

**Note.** ZCH2020: Zhang et al. (2020a); CDL2004: Ciardullo et al. (2004); MCM2000: Magrini et al. (2000); VGHC86: Viallefond et al. (1986); DMS93: Drissen et al. (1993); BCL74: Boulesteix et al. (1974); RVP2013: Relaño et al. (2013); LHK2017: Lin et al. (2017); B2011: Bresolin (2011); MVM2007: Magrini et al. (2007); CPS87: Courtes et al. (1987); LJD2011: Lagrois et al. (2011); CS82: Christian & Schommer (1982); MOH2006: Massey et al. (2006).

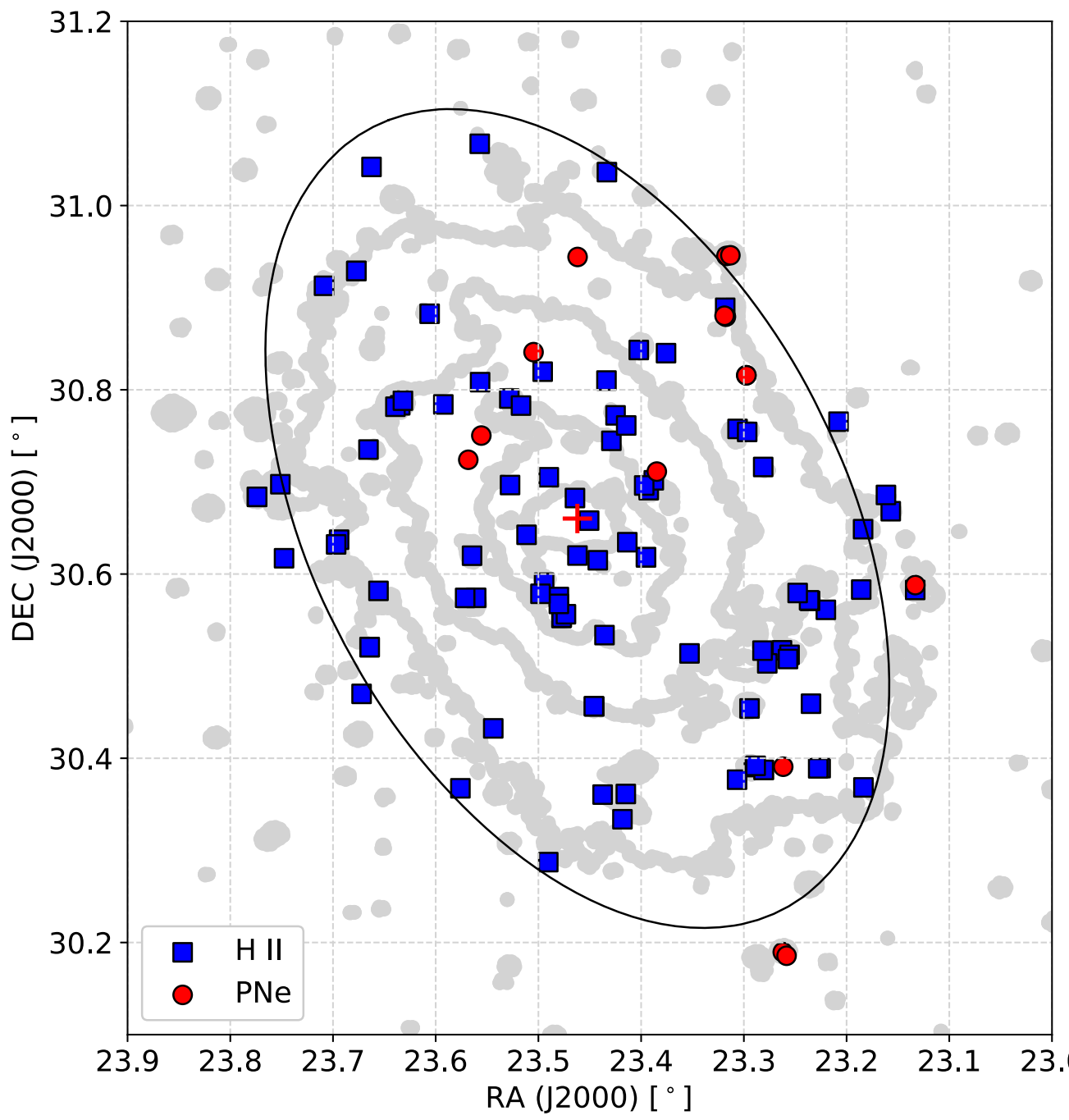

**Figure 2.** The distribution of 95 H II regions (squares) and 15 PNe (circles) in M33. The center of M33 is marked by a plus sign. The optical radius, $R_{25}$, is marked by a solid ellipse plotted with parameters (position angle and inclination) from Table 2.

**Table 2**
Parameters of M33

| Parameters (1) | Value (2) | Ref. (3) |
|---|---|---|
| Morphological type | SA(s)cd | 1 |
| R.A. (J2000) | $01^h33^m50\overset{s}{.}9$ | 2 |
| Decl. (J2000) | $+30°39'36\overset{''}{.}8$ | 2 |
| Distance (kpc) | 878 | 3 |
| Optical radius, $R_{25}$ | $28\overset{'}{.}2$ | 4 |
| Scale (pc arcmin$^{-1}$) | 255.4 | … |
| Position angle (major axis) | 23° | 5 |
| Inclination | 56° | 5 |
| Apparent magnitude, $m_V$ | 5.28 | 1 |
| Systemic velocity (km s$^{-1}$) | −179 | 1 |

**References.** (1) de Vaucouleurs et al. 1991; (2) Plucinsky et al. 2008; (3) Lin et al. 2017; (4) de Vaucouleurs 1959; (5) Zaritsky et al. 1989. Inclination is given in degrees, with 0° defined as face-on. Position angle of the major axis in degrees is measured north to east.

regions. Some objects are common with the catalog of Zhang et al. (2020a), for example, PNe: J013332.34+304240.3 (#4) and J013302.02+301108.2 (#15); H II regions: J013439.00+310231.0 (#21), J013301.32+303044.1 (#54), J013301.67+303027.9 (#65), J013313.57+302236.3 (#82), and J013442.54+305544.5 (#85). The nebula J013340.36+302001.6 (#84) was classified as an H II region in our study. However, this nebula is located very close to J013340.37+302001.6, which was classified as a new SNR candidate in Zhang et al. (2020a), who employed an additional classification based on Frew & Parker (2010). We classified J013402.76+303833.2 (#26) and J013350.88+303712.2 (#37) as H II regions, while they were classified as PNe in Ciardullo et al. (2004) and Magrini et al. (2000). Both regions are marked in Figure 1.

## 3. Method of Calculations

The adopted parameters of M33 and their references are presented in Table 2. The identification of the lines, fitting with a Gaussian profile, and measurement of line intensities were done employing FITYK software. Among 110 objects, 108 targets do not contain stellar contamination. The spectra of the remaining two H II regions (#46 and #110) are contaminated by broad emission lines from ionizing central stars. These lines were subtracted during the nebular flux measurements. All spectra were corrected by subtracting a globally fitted continuum.

The subsequent plasma diagnostics and interstellar reddening were treated using the code NEAT[3] (v.2.2) (Nebular Empirical Analysis Tool; Wesson et al. 2012). Line intensities were measured on scale H$\beta = 100$. The code NEAT corrects for interstellar reddening using the ratios of the H$\alpha$, H$\beta$, H$\gamma$, and H$\delta$ lines. Intrinsic ratios of the lines are first calculated assuming a temperature of 10,000 K and a density of 1000 cm$^{-3}$. The extinction law of Howarth (1983) was adopted for calculations of extinction coefficients $c$(H$\beta$). The subsequent extinction-corrected intensities were employed for calculation of the electron densities ($n_e$) and temperatures ($T_e$). Electronic density was derived from the intensities of the sulfur-line doublet [S II] $\lambda\lambda$6716, 6731 assuming a $T_e$ of 10,000 K. The scheme to calculate the densities and temperatures can be found in Wesson et al. (2012).

For 6 PNe and 21 H II regions from our sample, low- and high-excitation temperatures were calculated from the intensities of several emission-line ratios. Electron temperature $T_e$[N II] was obtained from ratio [N II] $\lambda5755/(\lambda6548 + \lambda6584)$, $T_e$[O II] from ratio $\lambda3727/(\lambda7320 + \lambda7330)$, and $T_e$[O III] from ratio $\lambda4363/(\lambda5007 + \lambda4959)$ (Osterbrock & Ferland 2006). For the remaining objects, a value of $T_e = 10{,}000$ K was assumed to calculate the chemical abundances. Ionic abundances are calculated from CELs using the temperature and density appropriate to their ionization potential. We used (where available) the lines of [O I] $\lambda\lambda$6300, 6364, [O II] $\lambda\lambda$3727, 7320, 7330, [O III] $\lambda\lambda$4363, 4959, 5007, [N II] $\lambda\lambda$5755, 6584, 6548, [Ne III] $\lambda$3869, [S II] $\lambda\lambda$4069, 4076, 6716, 6731, [S III] $\lambda\lambda$6312, 9069, [Ar III] $\lambda\lambda$7136, 7751, and [Cl III] $\lambda\lambda$5517, 5537. The ionic abundance is adopted as a weighted average of the abundances from each ion, if several lines from a given ion are presented.

For helium, lines of He I $\lambda\lambda$4471, 5876, 6678 were employed. The tabulated emissivities for helium were adopted from Porter et al. (2012). NEAT interpolates logarithmically in temperature and density between the tabulated values to determine the appropriate emissivity. The chemical abundances are determined by applying the ionization correction factors (ICFs) for unseen ions following Delgado-Inglada et al. (2014). The classical ICF was applied for nitrogen (N/O = N$^+$/O$^+$). For H II regions, oxygen abundances were determined with two strong-line diagnostics: O3N2 and N2 employing the calibration formulae from Marino et al. (2013). The validity intervals, $-1.1 < \mathrm{O3N2} < 1.9$ and $-1.6 < \mathrm{O3N2} < -0.2$, were considered.

The error in flux ±10% may led to error in temperature ±30 K. We estimated the uncertainties of the abundances by propagation of the errors in line fluxes ±10% and errors in temperature ±500 K. The calculation was done for H II region #45 with $T_e = 8333$ K and $n_e = 56.9$ cm$^{-3}$ (Table 4). The total

[3] http://www.nebulousresearch.org/codes/neat

**Table 3**
The Chemical Abundances and Radial Velocities of Nebulae

| No. | $c(H\beta)$ | $n_e$ $cm^{-3}$ | $T_e$[O III] K | $T_e$[N II] K | $T_e$[O II] K | O/H (N2) | O/H (O3N2) | O/H | N/H | Ne/H | S/H | Cl/H | Ar/H | He/H | $V$ (km $s^{-1}$) | $\sigma$ ($V_r$) (km $s^{-1}$) | $R$ (kpc) | $R/R_{25}$ |
|---|---|---|---|---|---|---|---|---|---|---|---|---|---|---|---|---|---|---|
| (1) | (2) | (3) | (4) | (5) | (6) | (7) | (8) | (9) | (10) | (11) | (12) | (13) | (14) | (15) | (16) | (17) | (18) | (19) |
| | | | | | | | | PNe | | | | | | | | | | |
| 1 | 0.729 | 1000.0 | … | … | … | … | … | 8.71 | … | … | 6.67 | … | … | … | −55 | 33 | 5.897 | 0.8188 |
| 2 | 0.132 | 148.3 | … | … | … | … | … | 8.68 | 8.01 | … | 6.96 | … | 5.94 | … | −101 | 27 | 2.267 | 0.3148 |
| 3 | 0.117 | 1.0 | 9656 | … | … | … | … | 8.44 | 7.77 | 7.76 | 6.51 | 5.15 | 6.14 | 10.76 | −72 | 25 | 7.404 | 1.0280 |
| 4 | 0.480 | 1.0 | … | … | … | … | … | 8.58 | 7.75 | … | 6.71 | … | … | 10.77 | −80 | 36 | 2.506 | 0.3479 |
| 5 | 0.286 | 46.7 | … | … | … | … | … | 8.63 | … | … | … | … | … | … | −104 | 0 | 2.926 | 0.4062 |
| 6 | 0.280 | 1000.0 | … | … | … | … | … | 8.31 | … | … | … | … | 5.7 | 10.36 | 94 | 0 | 8.030 | 1.1149 |
| 7 | 0.348 | 8.6 | 10,446 | … | … | … | … | 8.19 | 7.50 | 7.54 | 6.20 | … | 5.57 | 10.63 | −97 | 1 | 8.093 | 1.1237 |
| 8 | 0.079 | 51.8 | 10,129 | … | … | … | … | 8.20 | 7.90 | 7.60 | 6.56 | 4.66 | 6.11 | 10.84 | 1 | 7 | 6.383 | 0.8862 |
| 9 | 0.003 | 6.1 | … | … | … | … | … | 8.31 | 7.71 | … | 6.51 | … | … | … | 20 | 6 | 5.906 | 0.8200 |
| 10 | 0.073 | 300.8 | 10,568 | … | … | … | … | 8.17 | 7.84 | 7.60 | 6.81 | 4.93 | 6.14 | 10.87 | −10 | 7 | 6.360 | 0.8831 |
| 11 | 0.097 | 1.0 | … | … | … | … | … | 8.48 | 7.25 | 8.39 | 6.8 | … | 5.75 | 10.57 | 78 | 7 | 5.535 | 0.7685 |
| 12 | 0.490 | 280.4 | … | … | … | … | … | 8.49 | … | … | 6.65 | … | … | … | −50 | 9 | 2.530 | 0.3513 |
| 13 | 0.222 | 6.1 | 8914 | … | … | … | … | 8.69 | … | 8.48 | 6.53 | … | 6.15 | … | −79 | 7 | 5.034 | 0.6990 |
| 14 | 0.218 | 41.6 | 8451 | … | … | … | … | 8.58 | 7.28 | 7.95 | 6.52 | … | 5.73 | 10.49 | −45 | 5 | 7.470 | 1.0372 |
| 15 | 0.178 | 615.7 | … | … | … | … | … | 8.35 | 7.30 | … | 6.49 | … | 5.45 | 10.90 | 78 | 12 | 8.074 | 1.1210 |
| | | | | | | | | H II regions | | | | | | | | | | |
| 16 | 0.007 | 13.7 | … | … | … | 8.29 | 8.47 | 7.62 | 6.98 | … | 6.07 | … | … | 11.13 | −90 | 38 | 1.554 | 0.2157 |
| 17 | 0.001 | 31.5 | 8138 | … | … | 8.28 | 8.29 | 8.34 | 7.44 | 7.92 | 6.39 | … | 6.14 | 10.97 | −54 | 37 | 2.099 | 0.2915 |
| 18 | 0.003 | 31.5 | 8091 | … | … | 8.30 | 8.27 | 8.41 | 7.64 | 8.07 | 6.84 | … | 6.19 | 11.04 | −4 | 43 | 1.598 | 0.2218 |
| 19 | 0.115 | 260.1 | … | … | … | 8.19 | 8.32 | 7.66 | 7.00 | … | 6.15 | … | … | … | 114 | 61 | 4.770 | 0.6623 |
| 20 | 0.221 | … | … | … | … | 8.43 | 8.40 | 7.83 | 7.44 | … | 6.37 | … | … | … | 103 | 40 | 0.775 | 0.1076 |
| 21 | 0.171 | 77.2 | … | … | … | 8.23 | 8.23 | 8.19 | 6.86 | … | 6.21 | … | 5.58 | … | −96 | 20 | 6.520 | 0.9053 |
| 22 | 0.253 | 1.0 | 8690 | … | … | 8.27 | 8.33 | 8.29 | 7.11 | 7.90 | 6.51 | … | 5.98 | 10.81 | 55 | 19 | 5.310 | 0.7372 |
| 23 | 0.165 | 1.0 | … | … | … | 8.27 | 8.39 | 8.11 | 6.72 | … | 6.38 | … | 5.77 | 10.81 | −20 | 26 | 7.569 | 1.0508 |
| 24 | 0.069 | 1.0 | 7945 | … | … | 8.26 | 8.24 | 8.45 | 7.59 | 7.64 | 6.58 | … | 6.15 | 10.78 | −30 | 7 | 2.078 | 0.2885 |
| 25 | 0.638 | 1000.0 | … | … | … | 8.00 | 8.45 | 6.33 | 6.18 | 7.69 | … | … | … | 10.82 | −14 | 0 | 7.804 | 1.0835 |
| 26 | 0.07 | 1.0 | … | … | … | 8.47 | 8.48 | 7.79 | 7.29 | … | 6.48 | … | 5.55 | 10.70 | −31 | 18 | 1.447 | 0.2009 |
| 27 | 0.117 | 1.0 | … | … | … | 8.45 | 8.36 | 8.10 | 7.35 | … | 6.76 | … | … | … | −2 | 60 | 4.505 | 0.6255 |
| 28 | 0.208 | 1.0 | … | … | … | 8.46 | 8.41 | 7.84 | 7.53 | … | 6.63 | … | 5.82 | 11.02 | −9 | 18 | 0.297 | 0.0413 |
| 29 | 0.051 | 56.9 | … | … | … | 8.36 | 8.44 | 8.05 | 6.99 | … | 6.52 | … | 5.61 | 10.62 | −93 | 32 | 4.901 | 0.6805 |
| 30 | 0.142 | 1.0 | … | … | … | 8.36 | 8.30 | 8.15 | 7.06 | … | 6.44 | … | 5.34 | … | −84 | 45 | 4.990 | 0.6928 |
| 31 | 0.157 | 82.3 | … | … | … | 8.38 | 8.46 | 8.00 | 7.02 | … | 6.36 | … | … | … | −84 | 14 | 3.011 | 0.4181 |
| 32 | 0.474 | 321.1 | … | … | … | 8.39 | 8.32 | 7.80 | 6.57 | … | 6.01 | … | … | … | −120 | 60 | 6.508 | 0.9035 |
| 33 | 0.05 | 1.0 | … | … | … | 8.41 | 8.47 | 7.69 | 6.80 | … | 6.56 | … | 5.75 | 10.61 | 17 | 38 | 2.331 | 0.3236 |
| 34 | 0.004 | 1.0 | 8884 | … | … | 8.30 | 8.31 | 8.44 | 7.12 | 7.96 | 6.68 | … | 6.01 | 10.77 | 9 | 26 | 3.533 | 0.4905 |
| 35 | 0.067 | 13.7 | … | … | … | 8.28 | 8.26 | 8.07 | 7.32 | … | 6.33 | … | 5.74 | 10.79 | −3 | 24 | 1.786 | 0.2479 |
| 36 | 0.063 | 1.0 | … | … | … | 8.30 | 8.28 | 8.21 | 6.89 | … | 6.52 | … | … | … | −9 | 46 | 8.053 | 1.1181 |
| 37 | 0.081 | 49.3 | … | … | … | 8.49 | 8.51 | 8.22 | 7.02 | … | 6.52 | … | … | … | −25 | 37 | 0.732 | 0.1017 |
| 38 | 0.03 | 1.0 | … | … | … | 8.11 | 8.11 | 8.29 | 7.17 | … | 6.47 | … | 5.75 | 10.75 | 7 | 57 | 6.150 | 0.8539 |
| 39 | 0.213 | 82.3 | … | … | … | 8.16 | 8.21 | 8.00 | 7.33 | … | 6.47 | … | 5.76 | 10.80 | −1 | 69 | 5.138 | 0.7134 |
| 40 | 0.23 | 21.3 | 9091 | … | … | 8.11 | 8.13 | 8.29 | 7.54 | 7.46 | 6.60 | … | 5.86 | 10.77 | 15 | 52 | 5.216 | 0.7242 |
| 41 | 0.281 | 3.5 | … | … | … | 8.47 | 8.56 | 7.36 | 7.22 | … | 6.18 | … | 5.31 | 10.89 | −74 | 29 | 0.375 | 0.0521 |
| 42 | 0.26 | 163.6 | 8624 | … | … | 8.26 | 8.24 | 8.21 | 7.72 | 7.92 | 6.39 | … | 5.92 | 10.80 | −19 | 60 | 4.343 | 0.6030 |
| 43 | 0.335 | 1.0 | … | … | … | 8.27 | 8.43 | 7.55 | 6.66 | … | 5.96 | … | 4.97 | 10.09 | −16 | 37 | 6.722 | 0.9334 |
| 44 | 0.254 | 77.2 | … | … | … | 8.44 | 8.44 | 8.05 | 6.82 | … | 6.24 | … | 5.37 | 10.54 | 7 | 26 | 2.155 | 0.2992 |

**Table 3**
(Continued)

| No. | $c$(H$\beta$) | $n_e$ cm$^{-3}$ | $T_e$[O III] K | $T_e$[N II] K | $T_e$[O II] K | O/H (N2) | O/H (O3N2) | O/H | N/H | Ne/H | S/H | Cl/H | Ar/H | He/H | $V$ (km s$^{-1}$) | $\sigma$ (V$_r$) (km s$^{-1}$) | $R$ (kpc) | $R/R_{25}$ |
|---|---|---|---|---|---|---|---|---|---|---|---|---|---|---|---|---|---|---|
| (1) | (2) | (3) | (4) | (5) | (6) | (7) | (8) | (9) | (10) | (11) | (12) | (13) | (14) | (15) | (16) | (17) | (18) | (19) |
| 45 | 0.036 | 56.9 | 7922 | 8333 | ⋯ | 8.34 | 8.30 | 8.39 | 7.70 | 7.94 | 6.80 | 5.02 | 6.17 | 10.91 | 1 | 32 | 1.604 | 0.2228 |
| 46 | 0.088 | 112.8 | 8110 | 8917 | ⋯ | 8.31 | 8.26 | 8.36 | 7.73 | 7.72 | 6.81 | 4.94 | 6.13 | 10.79 | −47 | 44 | 4.111 | 0.5708 |
| 47 | 0.158 | 21.3 | 8098 | 9795 | ⋯ | 8.37 | 8.34 | 8.09 | 7.85 | 7.33 | 6.76 | 5.16 | 6.00 | 10.70 | −6 | 6 | 1.988 | 0.2760 |
| 48 | 0.111 | 41.6 | 7896 | 11,232 | ⋯ | 8.24 | 8.23 | 8.31 | 7.75 | 7.65 | 6.74 | 5.20 | 6.12 | 10.69 | 15 | 27 | 5.168 | 0.7175 |
| 49 | 0.057 | 16.2 | 7283 | 8889 | ⋯ | 8.36 | 8.34 | 8.31 | 7.72 | 7.66 | 6.83 | 5.14 | 6.13 | 10.74 | −41 | 7 | 2.229 | 0.3094 |
| 50 | 0.059 | 28.9 | 8492 | 9822 | ⋯ | 8.29 | 8.25 | 8.29 | 7.70 | 7.74 | 6.76 | 4.96 | 6.06 | 10.78 | −36 | 4 | 4.194 | 0.5823 |
| 51 | 0.168 | 34.0 | ⋯ | ⋯ | 6986 | 8.42 | 8.41 | 8.87 | 7.38 | ⋯ | 6.81 | ⋯ | 6.16 | 10.53 | 82 | 3 | 2.153 | 0.2990 |
| 52 | 0.157 | 1.0 | ⋯ | ⋯ | ⋯ | 8.24 | 8.21 | 7.95 | 6.85 | ⋯ | 6.43 | ⋯ | ⋯ | ⋯ | 83 | 15 | 4.470 | 0.6207 |
| 53 | 0.057 | 87.4 | ⋯ | ⋯ | ⋯ | 8.32 | 8.52 | 7.90 | 6.80 | ⋯ | 6.47 | ⋯ | 5.78 | ⋯ | 98 | 1 | 3.016 | 0.4188 |
| 54 | 0.009 | 1.0 | 10,434 | ⋯ | ⋯ | 8.38 | 8.30 | 8.23 | 7.17 | ⋯ | 6.56 | ⋯ | 5.83 | 10.72 | 79 | 1 | 4.948 | 0.6870 |
| 55 | 0.132 | 13.7 | ⋯ | ⋯ | ⋯ | 8.32 | 8.35 | 8.08 | 6.84 | ⋯ | 6.26 | ⋯ | 5.87 | 10.58 | 85 | 5 | 3.553 | 0.4933 |
| 56 | 0.197 | 1.0 | ⋯ | ⋯ | ⋯ | 8.39 | 8.39 | 8.07 | 7.15 | ⋯ | 6.84 | ⋯ | 5.83 | ⋯ | 69 | 1 | 1.791 | 0.2487 |
| 57 | 0.495 | 1.0 | ⋯ | ⋯ | ⋯ | 8.47 | 8.36 | 8.41 | 6.95 | ⋯ | 6.55 | ⋯ | ⋯ | ⋯ | −78 | 6 | 4.127 | 0.5730 |
| 58 | 0.055 | 41.6 | ⋯ | ⋯ | ⋯ | 8.38 | 8.41 | 8.17 | 6.95 | ⋯ | 6.83 | ⋯ | 5.56 | 10.73 | −5 | 4 | 3.508 | 0.4870 |
| 59 | 0.126 | 112.8 | ⋯ | ⋯ | ⋯ | 8.38 | 8.35 | 8.04 | 6.86 | ⋯ | 6.15 | ⋯ | ⋯ | ⋯ | 33 | 5 | 6.713 | 0.9321 |
| 60 | 0.059 | 204.2 | ⋯ | ⋯ | ⋯ | 8.45 | 8.41 | 8.14 | 7.12 | ⋯ | 6.89 | ⋯ | ⋯ | ⋯ | 58 | 6 | 5.030 | 0.6984 |
| 61 | 0.005 | 3.5 | ⋯ | ⋯ | ⋯ | 8.40 | 8.47 | 8.03 | 7.04 | ⋯ | 6.74 | ⋯ | ⋯ | ⋯ | 47 | 3 | 2.025 | 0.2811 |
| 62 | 0.012 | 1.0 | ⋯ | ⋯ | ⋯ | 8.28 | ⋯ | 8.20 | 6.67 | ⋯ | 6.01 | ⋯ | ⋯ | ⋯ | 51 | 9 | 7.079 | 0.9829 |
| 63 | 0.055 | 1.0 | ⋯ | ⋯ | ⋯ | 8.31 | 8.35 | 8.18 | 6.79 | ⋯ | 6.45 | ⋯ | 5.61 | 10.48 | 53 | 5 | 5.354 | 0.7433 |
| 64 | 0.066 | 1.0 | ⋯ | ⋯ | ⋯ | 8.38 | 8.43 | 8.04 | 6.97 | ⋯ | 6.92 | ⋯ | ⋯ | ⋯ | 27 | 5 | 5.823 | 0.8085 |
| 65 | 0.002 | 1.0 | ⋯ | ⋯ | ⋯ | 8.34 | 8.36 | 8.23 | 6.80 | ⋯ | 6.67 | ⋯ | 5.53 | 10.60 | 43 | 6 | 4.925 | 0.6838 |
| 66 | 0.042 | 8.6 | ⋯ | 8486 | ⋯ | 8.43 | 8.44 | 8.49 | 7.20 | 7.97 | 6.83 | ⋯ | 5.84 | 10.55 | −77 | 8 | 2.746 | 0.3813 |
| 67 | 0.006 | 1.0 | ⋯ | ⋯ | ⋯ | 8.31 | 8.35 | 8.25 | 6.81 | ⋯ | 6.54 | ⋯ | 5.63 | ⋯ | 12 | 10 | 5.769 | 0.8010 |
| 68 | 0.36 | 46.7 | ⋯ | ⋯ | ⋯ | 8.37 | 8.44 | 7.80 | 6.86 | ⋯ | 6.62 | ⋯ | ⋯ | ⋯ | 10 | 7 | 3.757 | 0.5217 |
| 69 | 0.073 | 1.0 | ⋯ | ⋯ | ⋯ | 8.40 | 8.53 | 8.13 | 6.72 | ⋯ | 6.71 | ⋯ | ⋯ | ⋯ | 43 | 4 | 5.462 | 0.7583 |
| 70 | 0.142 | 36.6 | ⋯ | ⋯ | ⋯ | 8.40 | 8.43 | 7.83 | 7.13 | ⋯ | 6.70 | ⋯ | 5.37 | 10.78 | 40 | 1 | 1.913 | 0.2656 |
| 71 | 0.191 | 128.0 | ⋯ | ⋯ | ⋯ | 8.22 | 8.18 | 8.20 | 6.97 | ⋯ | 6.19 | ⋯ | 5.57 | 10.68 | 66 | 8 | 7.083 | 0.9834 |
| 72 | 0.186 | 1.0 | ⋯ | ⋯ | ⋯ | 8.30 | 8.38 | 7.96 | 6.64 | ⋯ | 6.29 | ⋯ | 5.46 | 10.53 | 76 | 2 | 5.308 | 0.7370 |
| 73 | 0.172 | 143.3 | ⋯ | ⋯ | ⋯ | 8.29 | 8.35 | 7.43 | 6.54 | ⋯ | 5.90 | ⋯ | 5.45 | 10.54 | −76 | 10 | 6.133 | 0.8516 |
| 74 | 0.19 | 62.0 | ⋯ | ⋯ | ⋯ | 8.43 | 8.41 | 8.02 | 7.06 | ⋯ | 6.45 | ⋯ | 5.46 | ⋯ | −75 | 6 | 2.591 | 0.3597 |
| 75 | 0.204 | 1.0 | ⋯ | ⋯ | ⋯ | 8.38 | 8.38 | 8.12 | 6.74 | ⋯ | 6.12 | ⋯ | 5.49 | 10.39 | −4 | 8 | 2.316 | 0.3216 |
| 76 | 0.208 | 72.1 | ⋯ | ⋯ | ⋯ | 8.32 | 8.47 | 8.05 | 6.39 | ⋯ | 6.17 | ⋯ | ⋯ | ⋯ | −13 | 10 | 6.227 | 0.8646 |
| 77 | 0.16 | 1.0 | ⋯ | ⋯ | ⋯ | 8.46 | 8.44 | 7.89 | 7.09 | ⋯ | 6.24 | ⋯ | 5.47 | 10.59 | −59 | 5 | 0.820 | 0.1139 |
| 78 | 0.063 | 56.9 | ⋯ | ⋯ | ⋯ | 8.40 | 8.39 | 7.47 | 6.87 | ⋯ | 6.15 | ⋯ | 5.87 | 10.99 | −89 | 4 | 5.640 | 0.7830 |
| 79 | 0.05 | 31.5 | ⋯ | ⋯ | ⋯ | 8.46 | 8.39 | 8.15 | 7.26 | ⋯ | 6.81 | ⋯ | ⋯ | ⋯ | −61 | 4 | 3.203 | 0.4447 |
| 80 | 0.292 | 244.9 | ⋯ | ⋯ | ⋯ | 8.47 | 8.46 | 7.89 | 7.28 | ⋯ | 6.93 | ⋯ | ⋯ | ⋯ | −31 | 9 | 2.524 | 0.3504 |
| 81 | 0.001 | 31.5 | ⋯ | ⋯ | ⋯ | 8.33 | 8.40 | 8.13 | 6.85 | ⋯ | 6.58 | ⋯ | ⋯ | ⋯ | −49 | 8 | 4.539 | 0.6302 |
| 82 | 0.458 | 148.3 | ⋯ | ⋯ | ⋯ | 8.27 | 8.28 | 8.00 | 7.32 | ⋯ | 6.75 | ⋯ | 5.90 | ⋯ | 79 | 9 | 5.114 | 0.7101 |
| 83 | 0.014 | 36.6 | ⋯ | ⋯ | ⋯ | 8.34 | 8.51 | 7.83 | 6.86 | ⋯ | 6.84 | ⋯ | ⋯ | ⋯ | 42 | 5 | 5.408 | 0.7509 |
| 84 | 0.003 | 1.0 | ⋯ | ⋯ | ⋯ | 8.40 | 8.37 | 7.54 | 6.57 | ⋯ | 6.24 | ⋯ | ⋯ | ⋯ | 66 | 5 | 5.577 | 0.7743 |
| 85 | 0.17 | 1.0 | ⋯ | ⋯ | ⋯ | 8.40 | 8.39 | 7.95 | 7.06 | ⋯ | 6.45 | ⋯ | 5.72 | 10.69 | −84 | 7 | 5.654 | 0.7850 |
| 86 | 0.312 | 138.2 | ⋯ | ⋯ | ⋯ | 8.00 | 8.39 | 7.96 | 6.97 | ⋯ | 6.17 | ⋯ | 6.74 | ⋯ | −70 | 90 | 7.298 | 1.0132 |
| 87 | 0.145 | 16.2 | 9199 | ⋯ | ⋯ | 8.22 | 8.24 | 8.21 | 6.80 | 7.43 | 6.15 | ⋯ | 5.64 | 10.51 | −31 | 7 | 6.555 | 0.9102 |
| 88 | 0.242 | 67.0 | ⋯ | ⋯ | ⋯ | 8.25 | 8.29 | 8.09 | 6.59 | ⋯ | 6.13 | ⋯ | 5.43 | 10.45 | −91 | 7 | 6.575 | 0.9130 |
| 89 | 0.182 | 26.4 | 7948 | ⋯ | ⋯ | 8.24 | 8.24 | 8.44 | 7.29 | 7.68 | 6.48 | ⋯ | 5.83 | 10.58 | −54 | 5 | 4.111 | 0.5708 |
| 90 | 0.007 | 31.5 | ⋯ | ⋯ | ⋯ | 8.52 | 8.50 | 7.99 | 7.19 | ⋯ | 6.74 | ⋯ | ⋯ | ⋯ | −73 | 5 | 2.044 | 0.2838 |

**Table 3**
(Continued)

| No. | $c$(H$\beta$) | $n_e$ cm$^{-3}$ | $T_e$[O III] K | $T_e$[N II] K | $T_e$[O II] K | O/H (N2) | O/H (O3N2) | O/H | N/H | Ne/H | S/H | Cl/H | Ar/H | He/H | $V$ (km s$^{-1}$) | $\sigma$ (V$_r$) (km s$^{-1}$) | $R$ (kpc) | $R/R_{25}$ |
|---|---|---|---|---|---|---|---|---|---|---|---|---|---|---|---|---|---|---|
| (1) | (2) | (3) | (4) | (5) | (6) | (7) | (8) | (9) | (10) | (11) | (12) | (13) | (14) | (15) | (16) | (17) | (18) | (19) |
| 91 | 0.001 | 46.7 | ⋯ | ⋯ | ⋯ | 8.25 | 8.47 | 7.95 | 6.65 | ⋯ | 6.48 | ⋯ | ⋯ | ⋯ | −6 | 9 | 7.509 | 1.0425 |
| 92 | 0.116 | 3.5 | ⋯ | ⋯ | ⋯ | 8.25 | 8.26 | 8.17 | 7.02 | ⋯ | 6.53 | ⋯ | 5.81 | 10.73 | 80 | 6 | 5.143 | 0.7141 |
| 93 | 0.244 | 51.8 | ⋯ | ⋯ | ⋯ | 8.31 | 8.29 | 8.12 | 6.72 | ⋯ | 6.07 | ⋯ | 5.45 | 10.35 | −10 | 7 | 6.356 | 0.8824 |
| 94 | 0.207 | 1.0 | ⋯ | ⋯ | ⋯ | 8.41 | 8.39 | 7.93 | 7.06 | ⋯ | 6.46 | ⋯ | 5.58 | 10.50 | 33 | 4 | 2.050 | 0.2847 |
| 95 | 0.047 | 1.0 | ⋯ | ⋯ | ⋯ | 8.33 | 8.49 | 7.85 | 6.82 | ⋯ | 6.73 | ⋯ | 5.76 | ⋯ | 39 | 2 | 5.164 | 0.7170 |
| 96 | 0.247 | 1.0 | ⋯ | ⋯ | ⋯ | 8.24 | 8.21 | 8.22 | 6.77 | ⋯ | 6.37 | ⋯ | 5.19 | ⋯ | 15 | 8 | 7.139 | 0.9911 |
| 97 | 0.251 | 1.0 | ⋯ | ⋯ | ⋯ | 8.19 | 8.22 | 8.11 | 6.57 | 7.74 | 6.03 | ⋯ | 5.34 | 10.44 | 72 | 6 | 6.090 | 0.8456 |
| 98 | 0.137 | 1.0 | ⋯ | ⋯ | ⋯ | 8.24 | 8.31 | 8.12 | 6.30 | ⋯ | 5.94 | ⋯ | 5.25 | ⋯ | 29 | 7 | 7.968 | 1.1063 |
| 99 | 0.159 | 1.0 | ⋯ | ⋯ | ⋯ | 8.44 | 8.46 | 8.06 | 6.91 | ⋯ | 6.33 | ⋯ | 5.57 | 10.55 | −14 | 9 | 2.021 | 0.2807 |
| 100 | 0.117 | 1.0 | ⋯ | ⋯ | ⋯ | 8.23 | 8.23 | 8.18 | 7.01 | 7.89 | 6.51 | ⋯ | 5.74 | 10.74 | −38 | 5 | 6.955 | 0.9656 |
| 101 | 0.099 | 77.2 | 7432 | 9048 | ⋯ | 8.24 | 8.24 | 8.41 | 7.81 | 7.76 | 6.58 | 5.20 | 6.16 | 10.83 | 22 | 8 | 3.048 | 0.4232 |
| 102 | 0.083 | 1.0 | ⋯ | 8237 | ⋯ | 8.42 | ⋯ | 8.19 | 7.36 | 7.43 | 6.57 | ⋯ | 5.78 | 10.66 | 67 | 8 | 1.165 | 0.1617 |
| 103 | 0.135 | 11.2 | ⋯ | ⋯ | ⋯ | 8.35 | 8.56 | 7.48 | 6.77 | ⋯ | 6.16 | ⋯ | 5.17 | 10.62 | −4 | 3 | 2.542 | 0.3530 |
| 104 | 0.138 | 112.8 | 8190 | ⋯ | ⋯ | 8.22 | 8.22 | 8.25 | 7.78 | 7.53 | 6.55 | 5.09 | 6.07 | 10.83 | −34 | 6 | 4.059 | 0.5636 |
| 105 | 0.069 | 1.0 | ⋯ | ⋯ | ⋯ | 8.32 | 8.29 | 8.15 | 7.14 | ⋯ | 6.42 | ⋯ | 5.80 | 10.71 | −5 | 5 | 7.986 | 1.1088 |
| 106 | 0.232 | 74.7 | ⋯ | ⋯ | ⋯ | 8.24 | 8.42 | 8.13 | 6.88 | ⋯ | 6.44 | ⋯ | ⋯ | ⋯ | 106 | 4 | 7.104 | 0.9864 |
| 107 | 0.223 | 36.6 | ⋯ | ⋯ | ⋯ | 8.31 | 8.25 | 8.21 | 7.18 | 8.02 | 6.27 | ⋯ | 5.75 | 10.35 | 101 | 6 | 5.572 | 0.7736 |
| 108 | 0.138 | 158.5 | ⋯ | ⋯ | ⋯ | 8.31 | 8.31 | 8.00 | 7.27 | ⋯ | 6.53 | ⋯ | 5.87 | 10.51 | 60 | 6 | 6.726 | 0.9339 |
| 109 | 0.211 | 219.5 | ⋯ | ⋯ | ⋯ | 8.21 | 8.23 | 8.02 | 7.15 | 7.92 | 6.50 | ⋯ | ⋯ | 10.76 | 24 | 12 | 5.140 | 0.7136 |
| 110 | 0.012 | 1.0 | ⋯ | ⋯ | ⋯ | 8.26 | 8.24 | 8.15 | 6.92 | ⋯ | 5.90 | ⋯ | 5.41 | 10.71 | 36 | 5 | 7.960 | 1.1051 |

**Note.** A value of $T_e = 10{,}000$ K was assumed to calculate the ionic and total abundances, when $T_e$ is not available.

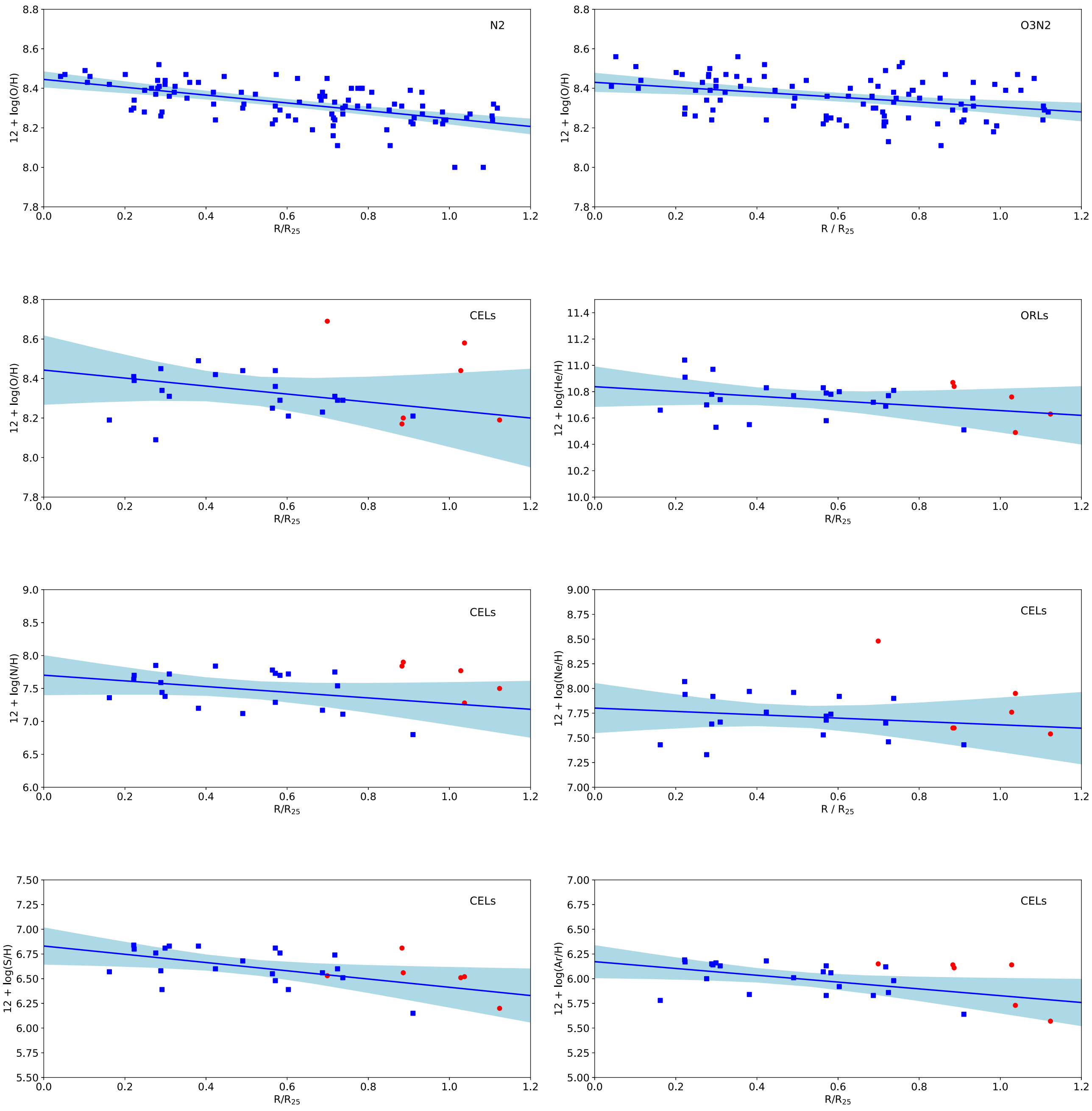

**Figure 3.** Radial metallicity gradients of O, He, N, Ne, S, and Ar for M33. Blue squares represent H II regions, while red circles represent PNe. The straight lines are the lines of best fits, and the shading represents the uncertainties around these fits. In the upper right corner of each panel the diagnostics are indicated. "N2" and "O3N2"—strong-line diagnostics. "CELs" and "ORLs" diagnostics—the chemical abundances were obtained with the $T_e$-sensitive method in 6 PNe and 21 H II regions.

uncertainties of oxygen abundances using N2 and O3N2 methods do not exceed $\sim\pm0.02$ dex. The uncertainties of the chemical abundances measured by the direct method can reach up to $\pm0.41$ dex.

We adopted the default atomic data in NEAT, that is, the transition probabilities of Zeippen (1982) for $O^+$ and Mendoza & Zeippen (1982) for $S^{2+}$, collision strengths of Pradhan (1976) for $O^+$ and Mendoza (1983) for $S^{2+}$, recombination data from Storey & Hummer (1995) for $H^+$ and Smits (1996) and Porter et al. (2012, 2013) for $He^+$, and transition probabilities and collision excitation rate coefficients from CHIANTI v. 7.0 (Landi et al. 2012) for the remaining species.

## 4. Results and Discussion

### 4.1. Chemical Gradients

We present the electron densities, temperatures, and total chemical abundances for targets in our sample in Table 3. The columns of Table 3 present (1) the identification number of the nebula; (2) the extinction coefficient $c(H\beta)$; (3) electronic density; (4) electron temperature from the detection of [O III] $\lambda$4363; (5) electron temperature from the detection of [N II] $\lambda$5755; (6) electron temperature from the detection of [O II] $\lambda$7320 and [O II] $\lambda$7330; (7) oxygen abundance obtained from strong-line index N2, which is defined as log ([N II]$\lambda$6583/H$\alpha$); (8) oxygen abundance obtained from strong-line index

**Table 4**
The Uncertainties in Abundances Calculated by Propagation of the Errors in the Line Fluxes and $T_e$

| Element | O/H | O/H | O/H | N/H | S/H | Ar/H | He/H |
|---|---|---|---|---|---|---|---|
| Method | N2 | O3N2 | CELs | CELs | CELs | CELs | ORLs |
| (1) | (2) | (3) | (4) | (5) | (6) | (7) | (8) |
| $\Delta F_\lambda = \pm 10\%$ | ±0.02 | ±0.02 | ±0.07 | ±0.09 | ±0.09 | ±0.08 | ±0.06 |
| $\Delta T_e = \pm 500$ K | 0 | 0 | ±0.20 | ±0.20 | ±0.30 | ±0.30 | ±0.40 |
| Total | ±0.02 | ±0.02 | ±0.21 | ±0.22 | ±0.31 | ±0.31 | ±0.41 |

**Table 5**
Chemical Radial Gradients in M33

| Element | Diagnostics | H II (dex $R_{25}^{-1}$) | H II (dex kpc$^{-1}$) |
|---|---|---|---|
| (1) | (2) | (3) | (4) |
| O | N2 | $-0.199^{+0.030}_{-0.030}$ | $-0.028^{+0.004}_{-0.004}$ |
| O | O3N2 | $-0.124^{+0.036}_{-0.036}$ | $-0.017^{+0.005}_{-0.005}$ |
| O | CELs | $-0.207^{+0.160}_{-0.174}$ | $-0.030^{+0.024}_{-0.024}$ |
| He | ORLs | $-0.179^{+0.145}_{-0.146}$ | $-0.026^{+0.020}_{-0.021}$ |
| N | CELs | $-0.431^{+0.282}_{-0.281}$ | $-0.070^{+0.043}_{-0.045}$ |
| Ne | CELs | $-0.171^{+0.234}_{-0.239}$ | $-0.027^{+0.035}_{-0.036}$ |
| S | CELs | $-0.417^{+0.174}_{-0.182}$ | $-0.060^{+0.025}_{-0.025}$ |
| Ar | CELs | $-0.340^{+0.156}_{-0.157}$ | $-0.050^{+0.022}_{-0.022}$ |

O3N2; (9) oxygen abundance obtained from CELs; (10) nitrogen abundance obtained from CELs; (11) neon abundance obtained from CELs; (12) sulfur abundance obtained from CELs; (13) chlorine abundance obtained from CELs; (14) argon abundance obtained from CELs; (15) helium abundance obtained from ORLs; (16) radial velocity with subtracted systemic velocity of M33, $-179$ km s$^{-1}$; (17) standard deviation for radial velocity; (18) $R$—galactocentric distance in kpc calculated in this study; and (19) $R/R_{25}$—galactocentric distance in the fraction of galactic optical radius.

The chemical abundances of O, He, N, Ne, S, and Ar as a function of galactocentric distance ($R/R_{25}$) are presented in Figure 3. The chemical abundances obtained with the $T_e$-sensitive method in 21 H II regions were used for metallicity gradient analysis. We fit the relation between abundances, 12 +log(M/H) (where M = O, He, Ne, S, Ar), and $R/R_{25}$ as a first-order polynomial, with Gaussian intrinsic scatter around this line with standard deviation in the 12+log(M/H) direction. This is accomplished using the Markov Chain Monte Carlo (MCMC) technique implemented in the EMCEE package of PYTHON (Foreman-Mackey et al. 2013). Throughout this study EMCEE is run using 50 walkers and 2000 steps. The chain for all the parameters is inspected to check for convergence. Once the chain has converged, the best fits are presented in Figure 3, where the shading areas represent the uncertainty around this fit. The uncertainty is taken as the difference between the 84th and 50th percentile as the upper limit and the difference between the 50th and 16th percentile as the lower limit. If the posterior distributions follow normal (Gaussian) distributions, then this equates to the 1$\sigma$ uncertainty. Final chemical gradients and 1$\sigma$ uncertainties in H II regions are summarized in Table 5 and illustrated in Figure 4. The spatial distribution of chemical abundances of O, He, Ne, S, Ar, and Cl (where available) in M33 is presented in Figure 5.

The H II regions are relatively young objects. They are large, massive (generally between $10^2$ and $10^3$ $M_\odot$) regions of gas that are ionized by the UV radiation emitted by recently formed OB-type massive stars, luminous blue variables (LBVs), or Wolf-Rayet (W-R) stars with typical effective temperatures between 25 and 50 kK. There are several known giant H II regions in M33: NGC 604, NGC 595, VGHC 2-22 (NGC 592), NGC 588 IC 131, and IC 142.

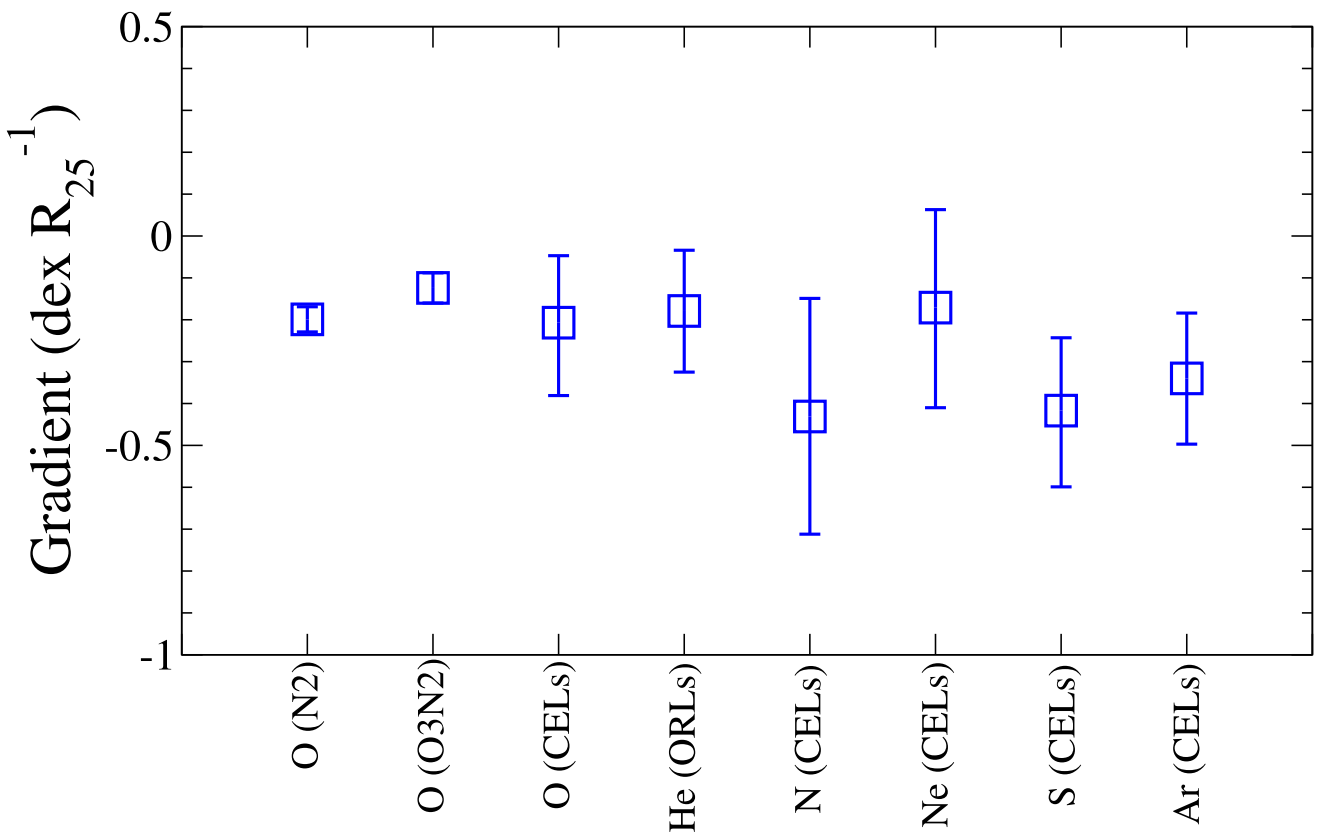

**Figure 4.** Radial metallicity gradients of O, He, N, Ne, S, and Ar for H II regions. On the horizontal axis, each element and the diagnostic method are shown.

A total of 55 H II regions were discovered for the first time in this study. The new H II regions are marked in Table 1. Some objects have combined spectra with both stellar and nebular characteristics. For example, J013441.32+302811.6 (#23) is an F-type star with nebular emission lines in spectra from an H II region. J013348.14+303928.6 (#28) is a G-type star with nebular emission lines in spectra from an H II region. J013316.36+305320.5 (#32) is an A-type star with nebular emission lines in spectra from an H II region. Although J013402.76+303833.2 (#26) was classified before as a PN (Ciardullo et al. 2004), we classified it as an H II region based on the BPT diagram (Figure 1).

Negative gradients were found for all presented elements in the sample of H II regions. We obtained the following O abundance gradients: $-0.199^{+0.030}_{-0.030}$ dex $R_{25}^{-1}$ at N2 diagnostics, $-0.124^{+0.036}_{-0.036}$ dex $R_{25}^{-1}$ at O3N2 diagnostics, and $-0.207^{+0.160}_{-0.174}$ at the direct abundance derivation.

PNe are expanding shells of the luminous gas expelled by dying stars of low and intermediate masses. They stem from objects that have lifetimes up to a gigayear. PNe are small and less massive ($\sim 10^{-1}$ $M_\odot$). After the envelope ejection, the remnant core of the star increases in temperature before the nuclear burning ceases and the star quickly fades, becoming a white dwarf. The high temperature of the central star ($\sim 25 \times 10^4$ K) causes the previously ejected material to be ionized, which becomes visible as a PN. There will be ionizing photons with enough energy to ionize high-excitation species, hence

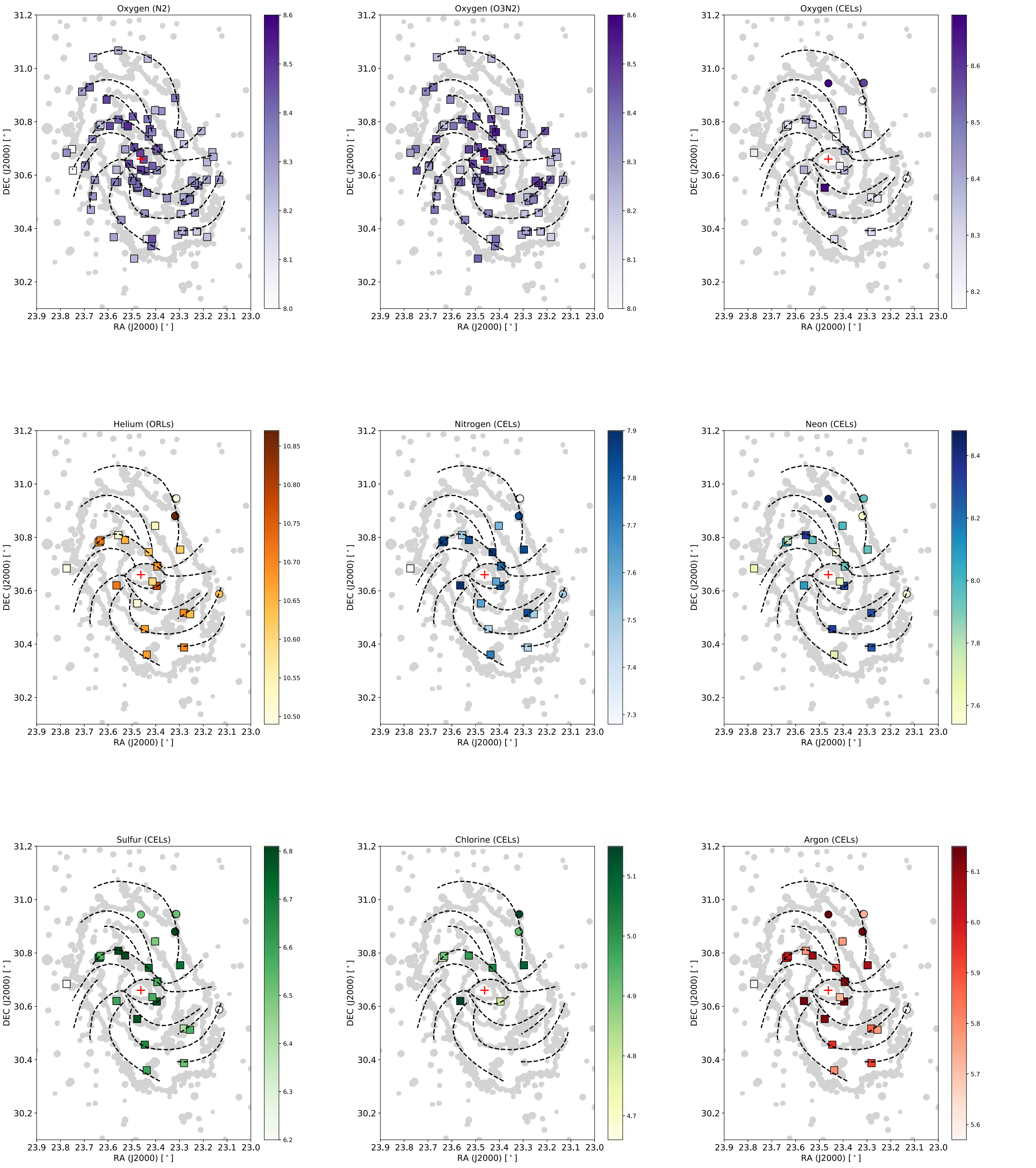

**Figure 5.** Spatial distribution of chemical elements in M33 is presented. Abundances of O, He, N, Ne, S, Cl, and Ar (where possible) in H II regions (squares) and PNe (circles) in M33 are shown by color bars. The diagnostics are indicated in brackets. Dashed lines show the spiral arms of the galaxy designed in this study.

producing qualitatively different spectra from that of H II regions.

Eight PNe from our sample were discovered and investigated for the first time (Table 1): #3, #6, #7, #8, #10, #11, #13, and #14. We spectroscopically confirmed that J013416.39 +304326.5 (#12) is a PN. We present the chemical abundances of O, He, N, Ne, S, and Ar (where available) obtained with the $T_e$-sensitive method in six PNe (Figure 3). We do not find strong deviation between the chemical compositions in six PNe and H II regions. The abundance gradient in PNe was not presented, because our PNe are located in the narrow range of galactocentric distances ($0.7 \leqslant R/R_{25} \leqslant 1.1$).

We have investigated the azimuthal distribution of some metals. The azimuthal O, N, and S distribution of H II regions in M33 is shown in Figure 6. The specific azimuths are the number of degrees, which increase counterclockwise from due north. The abundances of presented elements have nearly uniform azimuthal distribution within the uncertainty (rms scatter) of 0.02 dex for oxygen (N2 and O3N2 methods), 0.08 dex for nitrogen, and 0.18 dex for sulfur. These results are in line with previous studies (e.g., Bresolin 2011; Lin et al. 2017).

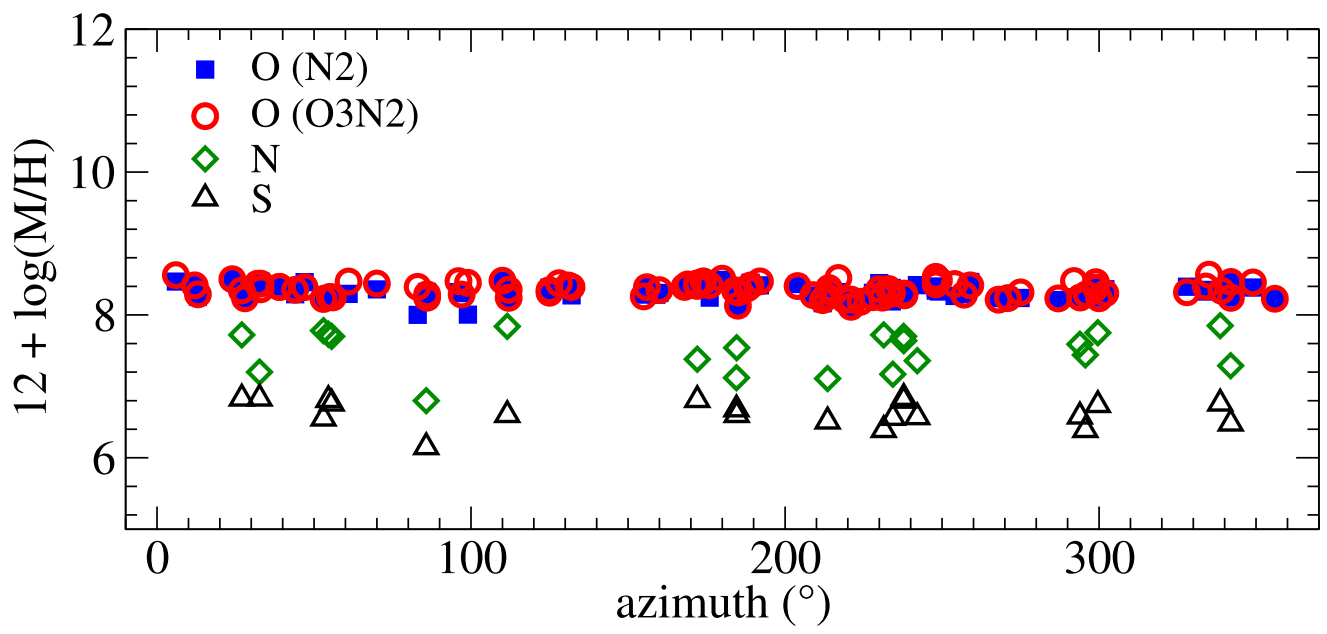

**Figure 6.** The azimuthal O, N, and S distribution of H II regions in M33. The diagnostic methods for oxygen are shown in brackets.

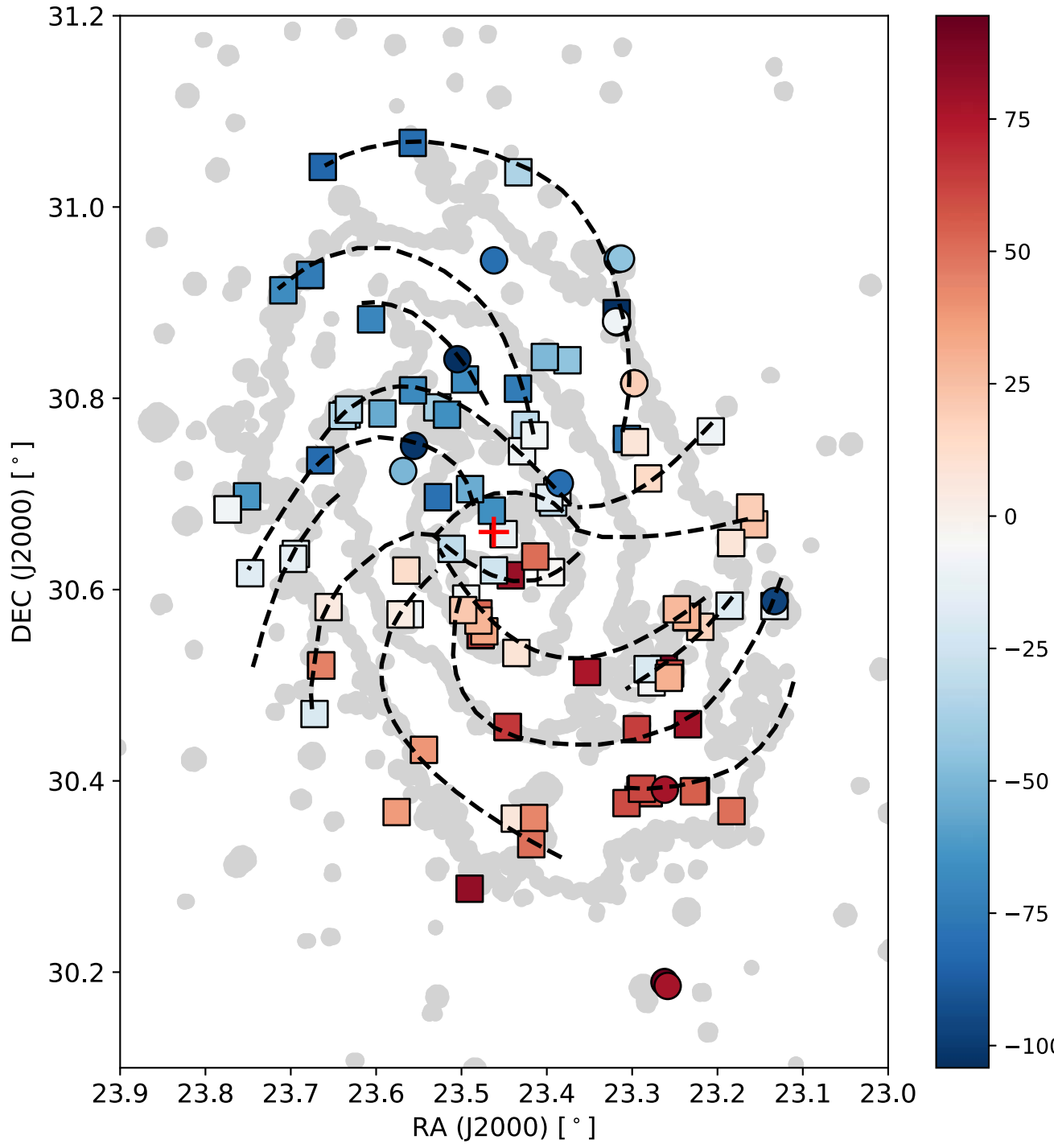

**Figure 7.** The distribution of the radial velocities of 110 objects in M33. The H II regions are marked by squares and PNe by circles. Dashed lines show the spiral arms of the galaxy designed in this study. The shadow region indicates the calculated radial velocities in km s$^{-1}$. Here the systemic velocity of 179 km s$^{-1}$ has been subtracted.

### 4.2. Radial Velocities

Radial velocities (Figure 7) were calculated with formula $Rv = c(\lambda/\lambda_0 - 1) + V_a$, where $c$ is speed of light, $\lambda$ are emission peaks of a line profile, $\lambda_0$ is laboratory wavelength, and $V_a$ is heliocentric correction. Heliocentric correction counting the orbital Earth rotation, daily rotation, and secular and Earth–Moon perturbations was calculated following the Stumpff (1980) prescriptions using the DECH package (Galazutdinov 1992). For each object (where possible), the wavelengths of the emission peaks of the H$\alpha$, H$\beta$, and [S II] $\lambda\lambda$6716, 6731 lines were considered. The error was calculated as standard deviation, $\sigma(V_r)$, from these lines. A sketch of spiral arms was adopted from Sandage & Humphreys (1980).

### 4.3. Spiral Structure of M33

H II regions are the sites of recent massive star formation, and they are good tracers for the galactic spiral structure. Here we present the grand-design pattern of the spiral structure of M33 by employing 519 H II regions. We collected 95 H II regions from this study, 396 nonoverlapping H II regions from Lin et al. (2017), and 28 nonoverlapping H II regions from Zhang et al. (2020a). Figure 8 presents the designed pattern of the spiral structure of M33. The sketches were drawn to follow the H II regions and trace the multiarm profile of M33. We identified 14 spiral arms in M33. Our grand-designed pattern is close to the one shown in Kam et al. (2015). However, more arms were noticed in this study compared to Kam et al. (2015).

### 4.4. NGC 604

NGC 604 is the largest star-forming region in M33 and the second most massive H II region in the Local Group and is thus an important laboratory for massive star formation. NGC 604 is also known as a depository of W-R stars. There are three objects in NGC 604 in our sample: J013432.29+304659.3 (#46), J013433.41+304653.8 (#50), and J013431.65 +304717.5 (#104). Below we comment on individual objects.

J013432.29 + 304659.3 (#46): The spectrum of this nebula contains both types of lines: narrow nebular and broad emission lines C III $\lambda$4650 and C IV $\lambda$5806 from an ionizing W-R star (Figure 10). Based on the relative strengths of these carbon lines, we classified this as a WC star. Possibly, this is a newly discovered WC star. However, we do not rule out that these carbon features may belong to an already-known WCE star classified by Massey & Johnson (1998) and studied in Neugent & Massey (2011) with its name J013432.32+304659.7, since the distance between J013432.32+304659.7 and J013432.29 +304659.3 is 2″.

J013433.41 + 304653.8 (#50): This object has no clear history of spectral classification in the literature prior to this work. We identified it as an H II region, and its spectrum is presented in Figure 9. The nebular part of its spectrum is similar to that of J013432.29+304659.3. Lines such as [O I], [O II], [O III], [N II], [Ne III], [S II], [S III], [Cl III], [Ar III], He I, He II, and hydrogen lines were measured. The chemical composition of this H II region was obtained for the first time. The chemical abundances of all investigated elements in both #46 and #50 are found to be similar.

J013431.65 + 304717.5 (#104): This object is located close to H II region LGGS J013431.69+304717.4.

### 4.5. NGC 595

NGC 595 is a large star-forming region in M33. There are three objects in NGC 595 in our sample: J013334.27+304136.7 (#17), J013334.22+304127.6 (#24), and J013335.28+304146.0 (#99). J013334.27+304136.7 (#17) was identified before as a W-R star in Neugent & Massey (2011). Both nebulae J013334.22 +304127.6 (#24) and J013335.28+304146.0 (#99) are investigated here for the first time.

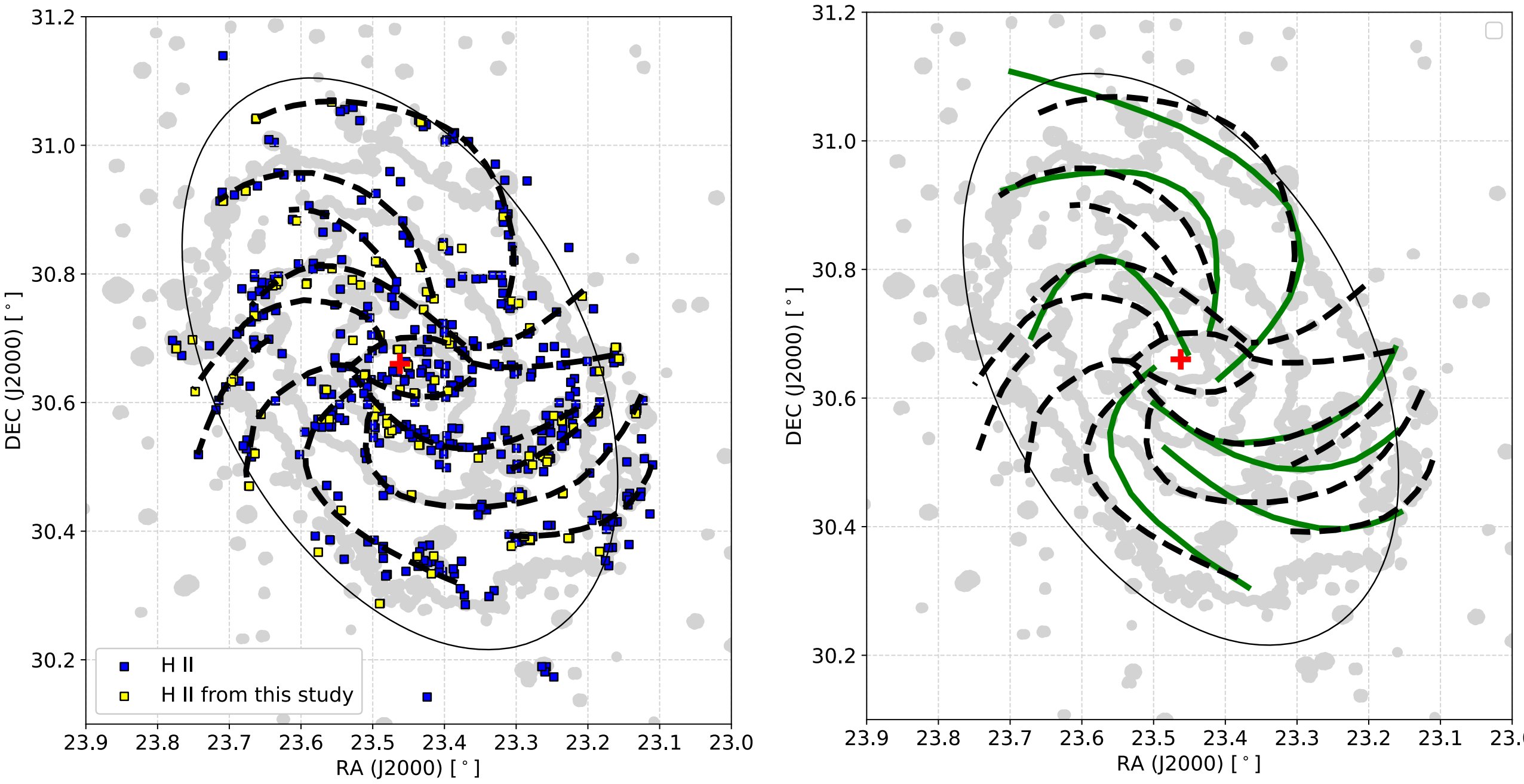

**Figure 8.** Left panel: the designed pattern of the spiral structure of M33 (black sketches). The dashed sketches are drawn to follow the H II regions. A total of 519 H II regions were employed, including this study (95 H II regions are marked by yellow squares), 396 nonoverlapping H II regions from Lin et al. (2017), and 28 nonoverlapping ones from Zhang et al. (2020a) (blue squares). Right panel: the comparison of our pattern of spiral structure (dashed sketches) with the one from Kam et al. (2015) (solid sketches).

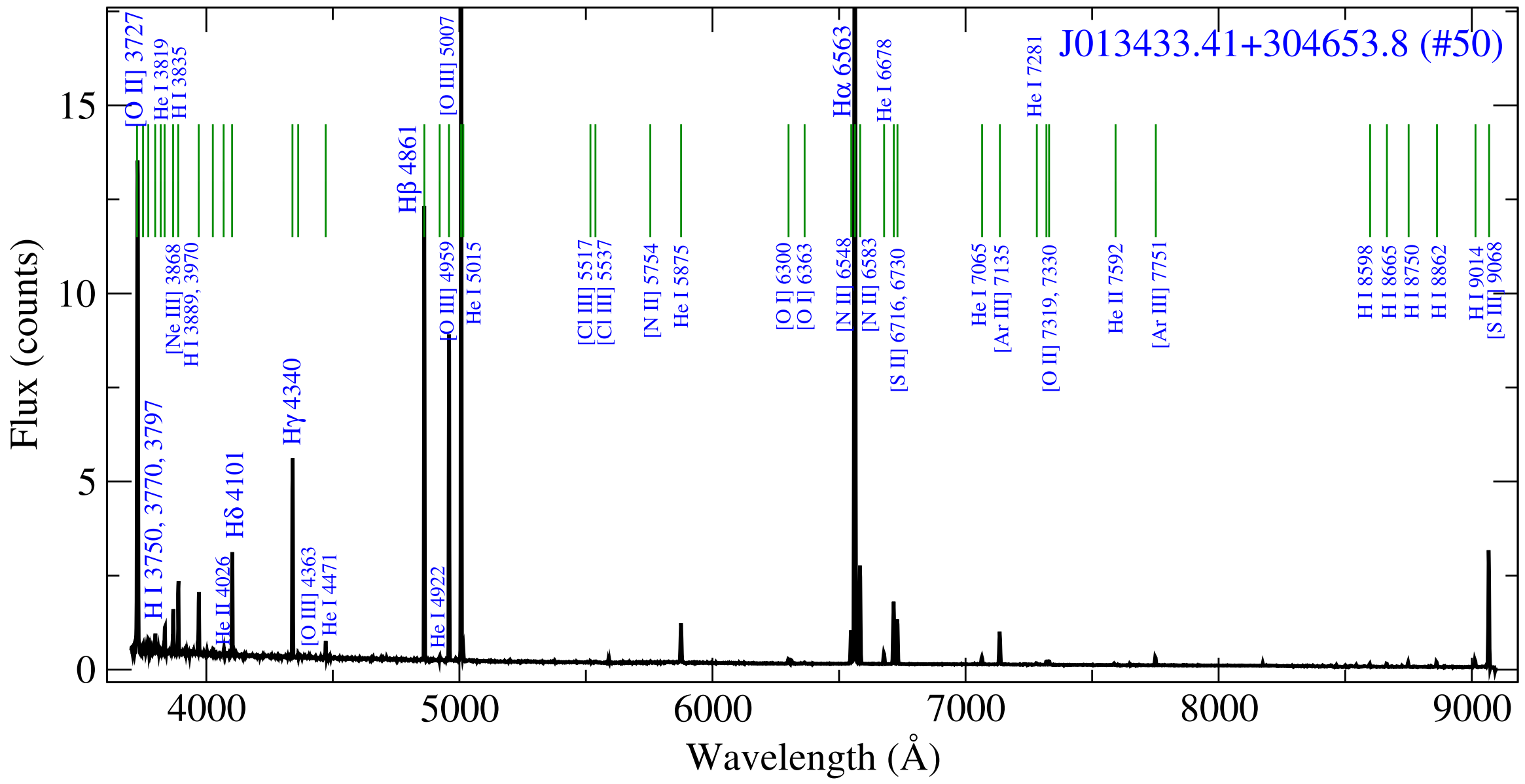

**Figure 9.** The spectrum of newly discovered H II region J013433.41+304653.8 (#50). All prominent emission lines are marked.

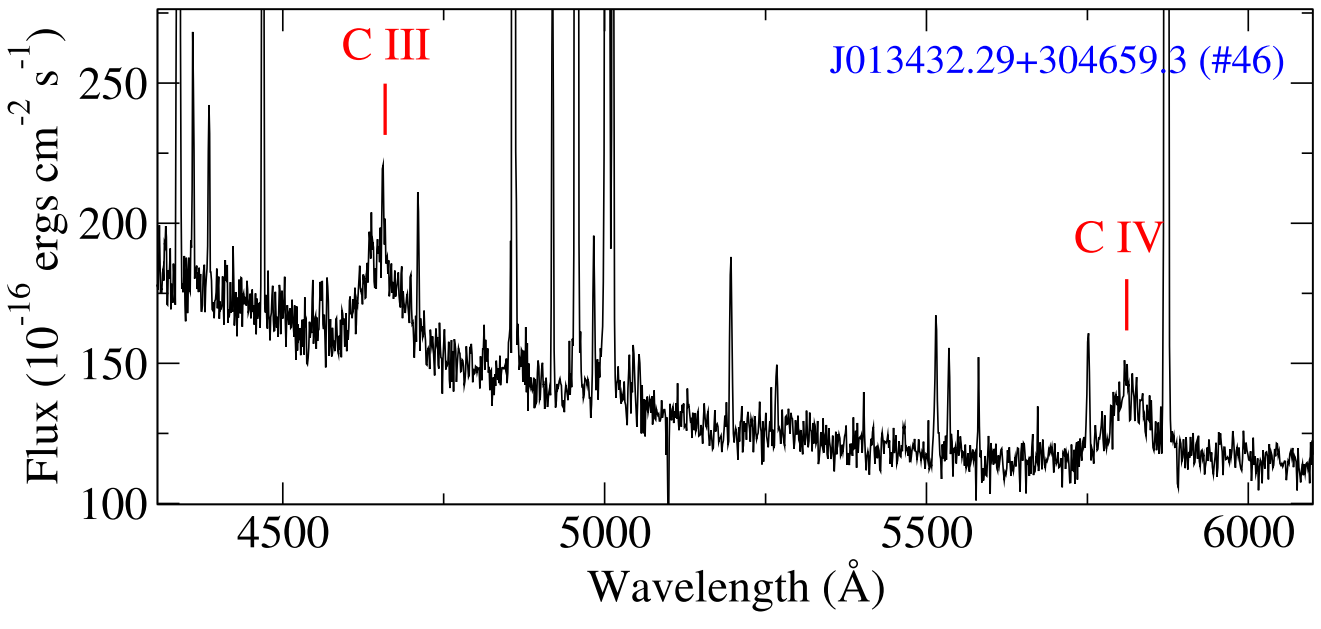

**Figure 10.** The spectrum of WC candidate star J013432.29+304659.3 (#46). The broad emission lines C III $\lambda$4650 and C IV $\lambda$5806 are marked. Plenty of the unmarked narrow emission lines are nebular.

### *4.6. New Transition W-R WN/C Star in M33?*

W-R stars are massive stars at the final evolutionary stage characterized by highly evolved surface elements and high mass-loss rates. Due to strong stellar winds, their spectra showed strong, broad emission lines of nuclear-processed material (Cassinelli & Hartmann 1975). W-R stars are divided into two types: WN (nitrogen rich) or WC (carbon rich), depending on the emission lines in their spectrum. Both types contain large amounts of helium with little to no hydrogen (Massey & Olsen 2003; Meynet & Maeder 2005; Meynet et al. 2011). In the spectra of some stars, the strong emission lines of both carbon and nitrogen can be detected, leading to an intermediate WN/C classification (Conti & Massey 1989; Crowther et al. 1995). WN/C stars are indeed considered to be at an intermediate evolutionary phase from the WN to WC stages. These transition stars are rarely observed in galaxies, because this transition period must be relatively short.

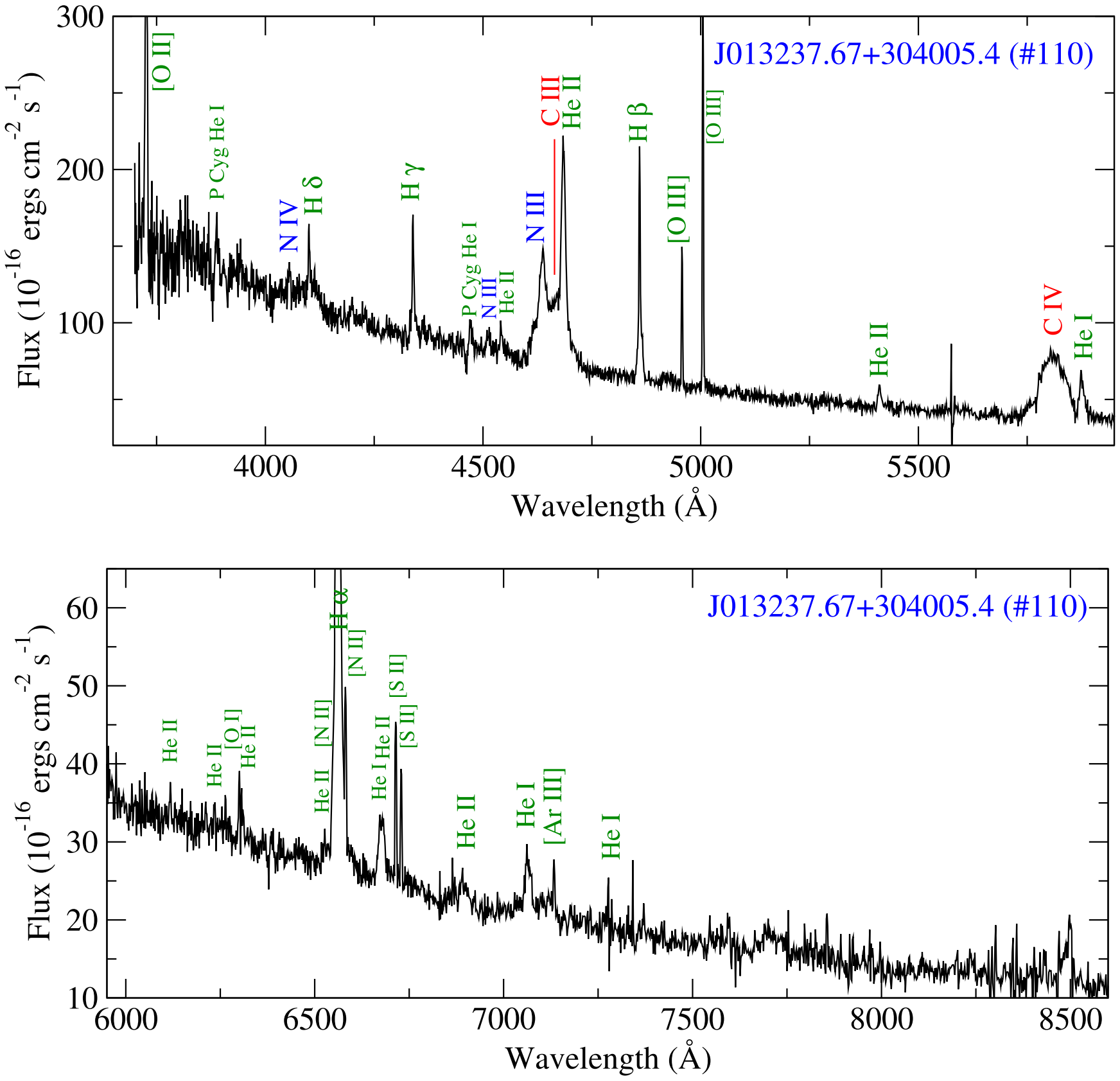

**Figure 11.** Top panel: the blue (3650–5950 Å) part of the spectrum of J013237.67+304005.4 (#110). The broad emission lines C III $\lambda$4650 and C IV $\lambda$5806 belong to the WC companion, while N III $\lambda\lambda$4634.42 and $\lambda\lambda$4511.15 belong to the WN companion. Bottom panel: the red (5950–9200 Å) part of the same star. Narrow emission lines are nebular.

The stars with WN/C characteristics were discovered as in M33 (Schild et al. 1990), as well as in other galaxies, e.g., in the Large Magellanic Cloud (Breysacher et al. 1999; Shenar et al. 2019; Hung et al. 2021), in the nearby spiral galaxy M81 (Gómez-González et al. 2016, 2020), and in our Galaxy (Zhang et al. 2020b). Recently, the first transition WN/C star was discovered in M31 (Shara et al. 2016).

Here we present a possible transition W-R WN/C star J013237.67+304005.4 (#110) in M33. The spectrum of #110 with line identification is presented in Figure 11. Except for narrow emission lines indicating evidence for circumstellar nebulae, there are plenty of broad emission lines. The broad emission lines C III $\lambda$4650 and C IV $\lambda$5806 allow us to classify the star as WC, and strong nitrogen lines such as N IV $\lambda$4058 and N III $\lambda\lambda$4634.42 and 4511.15 lead to WN classification; however, P Cygni profiles in He I $\lambda$3889 and He I $\lambda$4471 are main characteristics of Ofpe/WN9 "Slash" stars. These stars show properties intermediate between those of Of and WN stars and are believed to be a quiescent form of LBV (Massey et al. 1996). Our object has log EW(He II $\lambda$4686) = 1.56 and log EW(C IV $\lambda$5808) = 1.80, which are quite similar to those measured in other transition stars (Figure 5 in Conti & Massey 1989). According to the diagram of the log of EW of C IV $\lambda$5808 versus He II $\lambda$4686 from Conti & Massey (1989) and templates from Schild et al. (1990) and Massey & Johnson (1998), we classified J013237.67+304005.4 (#110) as WN8/C4-5.

To determine whether the carbon and nitrogen features arise in two stars or in a single object whose composition is intermediate between that of WN and WC stars, we measured radial velocities from the most emission lines that can be found in the spectra (Table 6). Each line was fitted with a Gaussian function, whose parameters are presented in Table 6. Some spectra are composite. For example, H$\alpha$ profiles contains two emission lines (broad and narrow). Their Gaussian fits are in the top panel of Figure 12. The region 4686 Å is even more complex, and four Gaussian functions are required to get the best fit of this region. Although the C III line has three components at 4647.42, 4650.25, and 4651.47 Å, we treated it as one line at 4649.445 Å. We found that both carbon lines C III $\lambda$4649.445 and C IV $\lambda$5806 are broader compared to other lines and the radial velocities are different. Using time domain data and three spectra of #110, obtained at the local times 23:50 (2018 October 9), 00:23, and 00:56 (2018 October 10), we found no wavelength shifts within this 1 hr period. However, the observational period is too short to completely rule out the binary system. We also cannot rule out that the spectrum of #110 is the spectrum of an Ofpe/WN9 "Slash" star, contaminated with WC features from a WC star, which is not gravitationally connected with #110.

## 5. Comparison with Previous Studies

Our results confirm the existence of the axisymmetric global metallicity distribution that is assumed in many previous

**Table 6**
Identified Lines and Their Radial Velocities in J013237.67+304005.4 (#110)

| Line Å (1) | Center (Å) (2) | RV* (km $s^{-1}$) (3) | HWHM (Å) (4) |
|---|---|---|---|
| Stellar | | | |
| N III 4199.83 | 4198.773 | 104 | 2.72 |
| N II 4056.9 | 4056.093 | 119 | 3.42 |
| Hγ 4340.47 | 4340.971 | 214 | 5.57 |
| N III 4514.86 | 4514.722 | 170 | 5.75 |
| He II 4541.59 | 4541.642 | 182 | 2.60 |
| N III 4634.14+4640.64 | 4636.214 | 103 | 6.52 |
| C III 4647.42+4650.25+4651.47 | 4645.084 | −102 | 28.60 |
| He II (narrow) 4685.68 | 4684.593 | 109 | 5.03 |
| He II (broad) 4685.68 | 4685.386 | 160 | 22.81 |
| Hβ 4861.33 | 4860.961 | 156 | 5.66 |
| He II 5411.52 | 5411.047 | 153 | 3.53 |
| C IV 5801.51+5812.14 | 5802.357 | −43 | 29.38 |
| He I 5875.66 | 5875.5 | 171 | 4.44 |
| Hα 6562.77 | 6561.786 | 134 | 10.09 |
| He I 6678.16+6683.2 | 6676.389 | −14 | 5.70 |
| He I 7065.25 | 7063.878 | 121 | 11.66 |
| Nebular | | | |
| [O II] 3727.00 | 3726.227 | 117 | 2.55 |
| Hδ 4101.74 | 4100.274 | 72 | 1.47 |
| Hγ 4340.47 | 4338.928 | 72 | 1.58 |
| Hβ 4861.33 | 4859.65 | 75 | 1.31 |
| [O III] 4958.91 | 4957.169 | 74 | 1.45 |
| [O III] 5006.84 | 5005.103 | 75 | 1.35 |
| Hα 6562.77 | 6560.88 | 93 | 2.08 |
| [N II] 6583.50 | 6582.046 | 113 | 2.61 |
| [S II] 6716.44 | 6714.627 | 98 | 2.02 |
| [S II] 6730.82 | 6728.777 | 88 | 2.14 |
| [Ar III] 7135.80 | 7133.382 | 77 | 2.67 |

**Note.** * Systemic velocity of M33, −179 km $s^{-1}$, has been subtracted.

studies (see, e.g., Monteverde et al. 1997; Urbaneja et al. 2005; Crockett et al. 2006; U et al. 2009; Magrini et al. 2009; Rosolowsky & Simon 2008; Magrini et al. 2010; Bresolin et al. 2010; Bresolin 2011; Toribio San Cipriano et al. 2016; Lin et al. 2017).

The biggest sample of 413 star-forming H II regions in M33 was investigated by Lin et al. (2017). They obtained O/H gradients fitted out to 1.1 $R_{25}$: $-0.17 \pm 0.03$, $-0.19 \pm 0.01$, and $-0.16 \pm 0.17$ dex $R_{25}^{-1}$, utilizing abundances from O3N2, N2 diagnostics, and a direct method, respectively. Our results are in agreement within the error bars with Lin et al. (2017) for all three methods. Toribio San Cipriano et al. (2016) performed a deep spectrophotometry of 11 H II regions in M33 and obtained oxygen gradients of $-0.33 \pm 0.13$ dex $R_{25}^{-1}$ (or $-0.045 \pm 0.13$ dex kpc$^{-1}$) using the ORLs and $-0.36 \pm 0.07$ dex $R_{25}^{-1}$ (or $-0.0495 \pm 0.07$ dex kpc$^{-1}$) from CELs. Their gradients are slightly higher in absolute value compared to our results for oxygen. Magrini et al. (2010) analyzed the spatial distribution of metals in M33 using a new sample and the literature data on H II regions and concluded that the metallicity gradient in the disk of M33 has a slope of $-0.27 \pm 0.009$ dex $R_{25}^{-1}$. A similar result was obtained by Bresolin (2011), who presented an oxygen gradient of $-0.29 \pm 0.07$ dex $R_{25}^{-1}$ from the analysis of CELs in 25 H II regions. Both of their gradients are slightly higher in absolute value compared to our evaluations for oxygen, but still in agreement within error bars. Rosolowsky & Simon (2008) presented a new determination of the metallicity gradient in M33, based on measurements of oxygen abundances using the $T_e$-sensitive emission-line [O III] λ4363 in 61 H II regions, and found a slope of $-0.19 \pm 0.08$ dex $R_{25}^{-1}$, which is close to our O/H gradient obtained with N2 diagnostics, $-0.199^{+0.030}_{-0.030}$ $R_{25}^{-1}$. By employing 13 H II regions in M33 and determining electron temperatures directly from the spectra using the [O III] λλ4959, 5007/λ4363 line ratio, Crockett et al. (2006) obtained an oxygen radial gradient of $-0.012 \pm 0.011$ dex kpc$^{-1}$ (or $-0.086 \pm 0.011$ dex $R_{25}^{-1}$), which is in line with our result of $-0.030^{+0.024}_{-0.024}$ dex kpc$^{-1}$ obtained with a similar method.

## 6. Conclusion

The study of the morphological and chemical structures of M33 can provide a constraint on models of the formation and evolution of the galaxies. We present the spiral structure and metallicity structure of M33 employing 110 nebulae in M33 from the LAMOST survey. A total of 63 newly identified nebulae (55 H II regions and 8 PNe) are investigated for the first time in this study. By employing commonly used BPT criteria, 110 nebulae in our sample were classified as 95 H II regions and 15 PNe. The plasma diagnostics; chemical composition of O, N, He, Ne, S, Cl, and Ar (where available); and interstellar reddening were treated using the code NEAT (v.2.2; Wesson et al. 2012). For 6 PNe and 21 H II regions, chemical abundances were obtained with the $T_e$-sensitive method, while $T_e = 10{,}000$ K was employed for the remaining objects. We obtained the following O abundance gradients: $-0.199^{+0.030}_{-0.030}$ dex $R_{25}^{-1}$ at N2 diagnostics (based on 95 H II regions), $-0.124^{+0.036}_{-0.036}$ dex $R_{25}^{-1}$ at O3N2 diagnostics (based on 93 H II regions), and $-0.207^{+0.160}_{-0.174}$ dex $R_{25}^{-1}$ at the direct abundance derivation (based on 21 H II region). The He, N, Ne, S, and Ar gradients resulted in slopes of $-0.179^{+0.145}_{-0.146}$, $-0.431^{+0.282}_{-0.281}$, $-0.171^{+0.234}_{-0.239}$, $-0.417^{+0.174}_{-0.182}$, and $-0.340^{+0.156}_{-0.157}$ dex $R_{25}^{-1}$, respectively, utilizing abundances from the $T_e$-sensitive method. The uncertainties in the chemical gradients were computed through Monte Carlo simulation. Less scatter (∼0.03 dex) was found in the oxygen gradient from the strong-line method with an N2 index. Negative axisymmetric metallicity gradients were obtained from all presented elements. This result confirms the existence of the global metallicity distribution that is assumed in numerous previous studies.

H II regions are good tracers of the sites of recent massive star formation. We use H II regions to probe the spiral structure of M33. Our sample of 95 H II regions was combined with 424 nonoverlapping H II regions known from the literature. Employing the sample with a total of 519 H II regions, we identified 14 spiral arms in M33. The reconstructed spiral structure of M33 is close to the one shown in Kam et al. (2015) from the distribution of the ionized gas.

During this study, we have found two spectra with a combination of the narrow nebulous emission lines and the broad emission lines that belonged to the mother star. Based on the spectral analysis, we report on possible discoveries: one WC star candidate J013432.29+304659.3 (#46) and one transition W-R WN/C candidate star J013237.67+304005.4 (#110).

All presented tables in this study, as well as measured fluxes, fluxes corrected for extinction, and ionic abundances in machine-readable format, are available via private request.

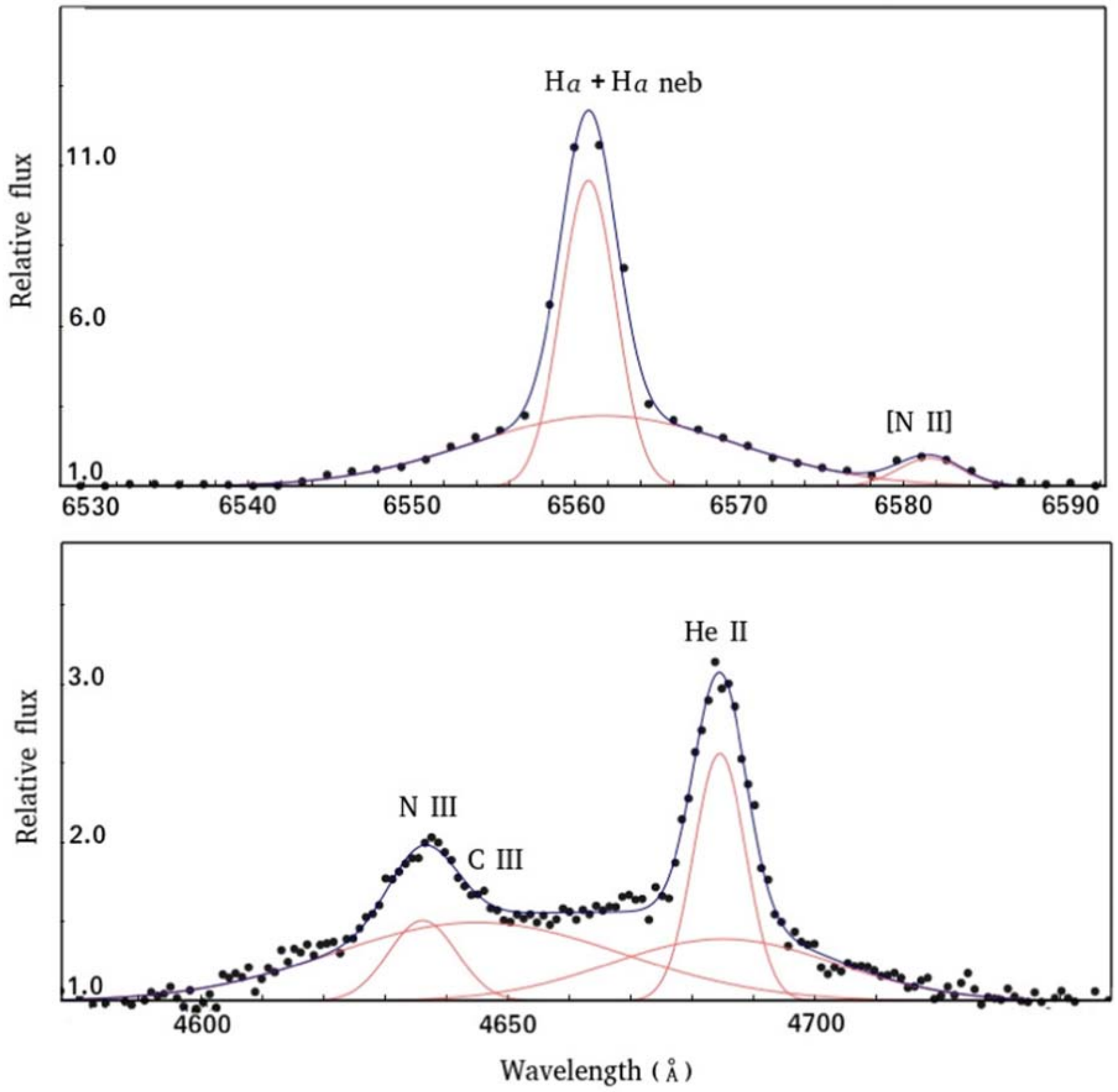

**Figure 12.** Gaussian functions and their sums (solid curves) resulting in the best fits of observations (black circles) of J013237.67+304005.4 (#110). Top panel: broad and narrow fits of the Hα profile. Bottom panel: N III λ4634.42, C III λλ4647.42, 4650.25, 4651.47, and He II λ4685.68 (broad and narrow). Parameters of Gaussian functions (center of line, HWHM) are listed in Table 6.

This study is supported by the National Natural Science Foundation of China under grant Nos. 11988101, 11890694, and 12050410265 and National Key R&D Program of China No. 2019YFA0405502. S.A. acknowledges support from the LAMOST Fellowship that is budgeted and administrated by the Chinese Academy of Sciences. We made use of the SIMBAD database, operated at CDS, Strasbourg, France.

Guoshoujing Telescope (the Large Sky Area Multi-Object Fiber Spectroscopic Telescope, LAMOST) is a National Major Scientific Project built by the Chinese Academy of Sciences. Funding for the project has been provided by the National Development and Reform Commission. LAMOST is operated and managed by the National Astronomical Observatories, Chinese Academy of Sciences.

*Software:* NEAT (v.2.2) (Wesson et al. 2012).

## ORCID iDs

Sofya Alexeeva https://orcid.org/0000-0002-8709-4665
Gang Zhao https://orcid.org/0000-0002-8980-945X

## References

Baldwin, J. A., Phillips, M. M., & Terlevich, R. 1981, PASP, 93, 5
Boulesteix, J., Courtes, G., Laval, A., Monnet, G., & Petit, H. 1974, A&A, 37, 33
Bresolin, F. 2011, ApJ, 730, 129
Bresolin, F. 2015, MmSAI, 86, 231
Bresolin, F., Stasińska, G., Vílchez, J. M., Simon, J. D., & Rosolowsky, E. 2010, MNRAS, 404, 1679
Breysacher, J., Azzopardi, M., & Testor, G. 1999, A&AS, 137, 117
Canto, J. 1981, in Investigating the Universe, ed. F. D. Kahn, 91 (Dordrecht: Reidel), 95
Cassinelli, J. P., & Hartmann, L. 1975, ApJ, 202, 718
Christian, C. A., & Schommer, R. A. 1982, ApJS, 49, 405
Ciardullo, R., Durrell, P. R., Laychak, M. B., et al. 2004, ApJ, 614, 167
Conti, P. S., & Massey, P. 1989, ApJ, 337, 251
Courtes, G., Petit, H., Sivan, J. P., Dodonov, S., & Petit, M. 1987, A&A, 174, 28
Crockett, N. R., Garnett, D. R., Massey, P., & Jacoby, G. 2006, ApJ, 637, 741
Crowther, P. A., Smith, L. J., & Willis, A. J. 1995, A&A, 304, 269
Cui, X.-Q., Zhao, Y.-H., Chu, Y.-Q., et al. 2012, RAA, 12, 1197
de Vaucouleurs, G. 1959, ApJ, 130, 728
de Vaucouleurs, G., de Vaucouleurs, A., Corwin, H. G., J., et al. 1991, S&T, 82, 621
Delgado-Inglada, G., Morisset, C., & Stasińska, G. 2014, MNRAS, 440, 536
Delgado-Inglada, G., Rodríguez, M., Peimbert, M., Stasińska, G., & Morisset, C. 2015, MNRAS, 449, 1797
Drissen, L., Moffat, A. F. J., & Shara, M. M. 1993, AJ, 105, 1400
Foreman-Mackey, D., Hogg, D. W., Lang, D., & Goodman, J. 2013, PASP, 125, 306
Frew, D. J., & Parker, Q. A. 2010, PASA, 27, 129
Galazutdinov, G. A. 1992, SAO Preprint SAO RAS, No. 92
Garnett, D. R., & Shields, G. A. 1987, ApJ, 317, 82
Gómez-González, V. M. A., Mayya, Y. D., & Rosa-González, D. 2016, MNRAS, 460, 1555
Gómez-González, V. M. A., Mayya, Y. D., Rosa-González, D., et al. 2020, MNRAS, 493, 3879
Howarth, I. D. 1983, MNRAS, 203, 301
Hubble, E. P. 1926, ApJ, 63, 236
Hung, C. S., Ou, P.-S., Chu, Y.-H., Gruendl, R. A., & Li, C.-J. 2021, ApJS, 252, 21
Ivanov, G. R., & Kunchev, P. Z. 1985, Ap&SS, 116, 341
Kam, Z. S., Carignan, C., Chemin, L., Amram, P., & Epinat, B. 2015, MNRAS, 449, 4048

Kang, X., Chang, R., Yin, J., et al. 2012, MNRAS, 426, 1455
Kauffmann, G., Heckman, T. M., Tremonti, C., et al. 2003, MNRAS, 346, 1055
Kniazev, A. Y., Pustilnik, S. A., & Zucker, D. B. 2008, MNRAS, 384, 1045
Kwitter, K. B., & Aller, L. H. 1981, MNRAS, 195, 939
Lagrois, D., Joncas, G., Drissen, L., & Arsenault, R. 2011, MNRAS, 413, 705
Landi, E., Del Zanna, G., Young, P. R., Dere, K. P., & Mason, H. E. 2012, ApJ, 744, 99
Li, H.-B., & Henning, T. 2011, Natur, 479, 499
Lin, Z., Hu, N., Kong, X., et al. 2017, ApJ, 842, 97
Magrini, L., Corradi, R. L. M., Mampaso, A., & Perinotto, M. 2000, A&A, 355, 713
Magrini, L., Stanghellini, L., Corbelli, E., Galli, D., & Villaver, E. 2010, A&A, 512, A63
Magrini, L., Stanghellini, L., & Villaver, E. 2009, ApJ, 696, 729
Magrini, L., Vílchez, J. M., Mampaso, A., Corradi, R. L. M., & Leisy, P. 2007, A&A, 470, 865
Marino, R. A., Rosales-Ortega, F. F., Sánchez, S. F., et al. 2013, A&A, 559, A114
Massey, P., Bianchi, L., Hutchings, J. B., & Stecher, T. P. 1996, ApJ, 469, 629
Massey, P., & Johnson, O. 1998, ApJ, 505, 793
Massey, P., & Olsen, K. A. G. 2003, AJ, 126, 2867
Massey, P., Olsen, K. A. G., Hodge, P. W., et al. 2006, AJ, 131, 2478
Mendoza, C. 1983, in Planetary Nebulae, ed. L. H. Aller, 103 (Dordrecht: Reidel), 143
Mendoza, C., & Zeippen, C. J. 1982, MNRAS, 198, 127
Meynet, G., Georgy, C., Hirschi, R., et al. 2011, BSRSL, 80, 266
Meynet, G., & Maeder, A. 2005, A&A, 429, 581
Mollá, M., Díaz, Á. I., Cavichia, O., et al. 2019, MNRAS, 482, 3071
Monteverde, M. I., Herrero, A., Lennon, D. J., & Kudritzki, R. P. 1997, ApJL, 474, L107
Neugent, K. F., & Massey, P. 2011, ApJ, 733, 123
Osterbrock, D. E., & Ferland, G. J. 2006, Astrophysics of Gaseous Nebulae and Active Galactic Nuclei (Sausalito, CA: Univ. Science Books)
Pagel, B. E. J., Edmunds, M. G., Blackwell, D. E., Chun, M. S., & Smith, G. 1979, MNRAS, 189, 95
Pease, F. G. 1915, PASP, 27, 239
Peña, M., & Flores-Durán, S. N. 2019, RMxAA, 55, 255
Plucinsky, P. P., Williams, B., Long, K. S., et al. 2008, ApJS, 174, 366
Porter, R. L., Ferland, G. J., Storey, P. J., & Detisch, M. J. 2012, MNRAS, 425, L28
Porter, R. L., Ferland, G. J., Storey, P. J., & Detisch, M. J. 2013, MNRAS, 433, L89
Pradhan, A. K. 1976, MNRAS, 177, 31
Relaño, M., Verley, S., Pérez, I., et al. 2013, A&A, 552, A140
Riesgo, H., & López, J. A. 2006, RMxAA, 42, 47
Rosolowsky, E., & Simon, J. D. 2008, ApJ, 675, 1213
Sabbadin, F., Minello, S., & Bianchini, A. 1977, A&A, 60, 147
Sandage, A., & Humphreys, R. M. 1980, ApJL, 236, L1
Sanders, N. E., Caldwell, N., McDowell, J., & Harding, P. 2012, ApJ, 758, 133
Schild, H., Smith, L. J., & Willis, A. J. 1990, A&A, 237, 169
Searle, L. 1971, ApJ, 168, 327
Shara, M. M., Mikołajewska, J., Caldwell, N., et al. 2016, MNRAS, 455, 3453
Shenar, T., Sablowski, D. P., Hainich, R., et al. 2019, A&A, 627, A151
Slipher, V. M. 1915, PA, 23, 21
Smits, D. P. 1996, MNRAS, 278, 683
Storey, P. J., & Hummer, D. G. 1995, MNRAS, 272, 41
Stumpff, P. 1980, A&AS, 41, 1
Toribio San Cipriano, L., García-Rojas, J., Esteban, C., Bresolin, F., & Peimbert, M. 2016, MNRAS, 458, 1866
U, V., Urbaneja, M. A., Kudritzki, R.-P., et al. 2009, ApJ, 704, 1120
Urbaneja, M. A., Herrero, A., Kudritzki, R. P., et al. 2005, ApJ, 635, 311
Viallefond, F., Goss, W. M., van der Hulst, J. M., & Crane, P. C. 1986, A&AS, 64, 237
Wesson, R., Stock, D. J., & Scicluna, P. 2012, MNRAS, 422, 3516
Zaritsky, D., Elston, R., & Hill, J. M. 1989, AJ, 97, 97
Zeippen, C. J. 1982, MNRAS, 198, 111
Zhang, M., Chen, B.-Q., Huo, Z.-Y., et al. 2020a, RAA, 20, 097
Zhang, W., Todt, H., Wu, H., et al. 2020b, ApJ, 902, 62
Zhao, G., Chen, Y.-Q., Shi, J.-R., et al. 2006, ChJA&A, 6, 265
Zhao, G., Zhao, Y.-H., Chu, Y.-Q., Jing, Y.-P., & Deng, L.-C. 2012, RAA, 12, 723